\documentclass[apj]{emulateapj}

\slugcomment{Accepted for publication in ApJ}
\shorttitle{Early Cosmological HII regions}
\shortauthors{Kitayama et al.}

\newcommand{\lbkt}[1]{\left[#1\right]}

\newcommand{\sbkt}[1]{\left(#1\right)}

\newcommand{\msun}{\,{\rm M}_{\odot}}
\newcommand{\HII}{H{\sc ii} }

\begin{document}

\title{The structure and evolution of early cosmological HII regions}

\author{Tetsu Kitayama}
\affil{Department of Physics, Toho University, Funabashi,
Chiba 274-8510 Japan;  kitayama@ph.sci.toho-u.ac.jp}
\author{Naoki Yoshida}
\affil{Department of Physics, Nagoya University, Chikusa-ku, Nagoya
464-8602, Japan}
\affil{National Astronomical Observatory of Japan, Mitaka,
Tokyo 186-8588, Japan}
\author{Hajime Susa}
\affil{Department of Physics, Rikkyo University, Toshimaku, Tokyo
171-8501, Japan}
\and
\author{Masayuki Umemura}
\affil{Center for Computational Physics, University of
Tsukuba, Tsukuba 305-8577, Japan}

\begin{abstract}
We study the formation and evolution of \HII regions around the first
stars formed at redshifts $z=10-30$.  We use a one-dimensional
Lagrangian hydrodynamics code which self-consistently incorporates
radiative transfer and non-equilibrium primordial gas chemistry.  The
star-forming region is defined as a spherical dense molecular gas cloud
with a Population III star embedded at the center.  We explore a large
parameter space by considering, as plausible early star-forming sites,
dark matter halos of mass $M_{\rm halo} = 10^5-10^8 \msun$, gas density
profiles with a power-law index $w=1.5-2.25$, and metal-free stars of
mass $M_{\rm star}=25 - 500 \msun$.  The formation of the \HII region is
characterized by initial slow expansion of a weak D-type ionization
front near the center, followed by rapid propagation of an R-type front
throughout the outer gas envelope.  We find that the transition between
the two front types is indeed a critical condition for the complete
ionization of halos of cosmological interest.  In small mass ($\la 10^6
\msun$) halos, the transition takes place within a few $10^5$ yr,
yielding high escape fractions ($>80\%$) of both ionizing and
photodissociating photons. The gas is effectively evacuated by a
supersonic shock, with the mean density within the halo decreasing to
$\la 1 {\rm cm}^{-3}$ in a few million years.  In larger mass ($\ga 10^7
\msun$) halos, the ionization front remains to be of D-type over the
lifetime of the massive star, the \HII region is confined well inside
the virial radius, and the escape fractions are essentially zero. We
derive an analytic formula, that reproduces well the results of our
simulations, for the critical halo mass below which the gas is
completely ionized.  We discuss immediate implications of the present
results for the star formation history and early reionization of the
Universe.
\end{abstract}


\section{Introduction}
The emergence of the first generation stars have significant impacts on
the thermal state of the inter-galactic medium (IGM) in the early
universe.  The initially neutral cosmic gas is photoionized and
photoheated by radiation from the first stars. This so-called radiative
feedback from the first stars is of considerable cosmological interest.
It can not only self-regulate the first star formation, but also
affect the formation and evolution of proto-galaxies.

Theories based on Cold Dark Matter (CDM) predict that the first
cosmological objects form when small mass ($\sim 10^6 \msun$) dark halos
assemble at redshifts $z=20-30$ (Couchman \& Rees 1986; Tegmark et
al. 1997; Abel et al. 1998; Yoshida et al. 2003a).  Stars formed in
these pre-galactic objects may have substantially contributed to cosmic
reionization.  Recent measurement of the large Thomson optical depth by
the WMAP satellite indicates that the universe was reionized at an epoch
as early as $z\sim 17$ (Kogut et al. 2003; Spergel et al. 2003),
supporting the above notion that the first generation stars formed at $z
\ga 20$.  Theoretical studies of the processes of cosmic reionization
by stellar sources suggest that reionization proceeds first by the
formation of individual \HII regions around radiation sources
(galaxies), and then by percolation of the growing \HII bubbles (Gnedin
\& Ostriker 1997; Ricotti et al. 2002; Sokasian et al. 2004).  The shape
and the extension of the individual \HII regions critically determine
the global topology of the ionized regions in a cosmological volume at
different epochs during reionization.

Studies on the formation of \HII regions in dense gas clouds date back
to the seminal work by Str\"omgren (1939).  Since then the structure of
\HII regions and the interaction with the surrounding medium have been
extensively studied (see Yorke 1986 for a review).  An important advance
has been made by Franco, Tenorio-Tagle \& Bodenheimer (1990) who
considered realistic conditions in which the initial gas density profile
is given by a power-law. Franco et al. showed that a hydrodynamic shock
effectively sweeps the ambient medium into a thin shell and the gas
density profile is significantly modified from the initial power-law
shape. Shu et al.  (2002) obtained the self-similar solutions for the
expansion of self-gravitating \HII regions. On cosmological backgrounds,
Shapiro \& Giroux (1987) studied the evolution of the cosmological
Str\"omgren sphere around luminous quasars in an expanding
universe. Ricotti \& Shull (2000) computed the UV photon consumption in
small mass halos to estimate the photon escape fraction.  Their
calculations, however, do not include the hydrodynamic response of the
photoheated gas and are not adequate to address the dynamical evolution
of \HII regions.  More recently, Whalen, Abel \& Norman (2004) carried
out a numerical simulation of ionization front (I-front) propagation
starting from a realistic initial density profile for the first
star-forming cloud.  Whalen et al. found that the escape fraction of
ionizing photons is close to unity in a particular case with halo mass
$7 \times 10^5 \msun$ at $z \sim 20$. It is by no means trivial if such
high escape fraction is achieved for different masses and redshifts.

In the present paper, we study the evolution of \HII regions around
Population III stars for a wide range of halo mass, redshift, and gas
density profile. Specifically, we consider an ``inside-out'' situation,
where the central source ionizes the surrounding gas and drives a
supersonic gas flow outward.  While the overall configuration of our
simulations may appear similar to those in conventional studies on \HII
regions, we have several critical ingredients relevant to the case of
early generation star-formation; 1) gravitational force exerted by the
host dark matter halo, 2) radiation force by the central star, 3)
radiative transfer of UV photons, 4) chemical reactions including
formation and destruction of hydrogen molecules, and 5) radiative
cooling and heating processes, including inverse-Compton cooling
particularly important at $z>10$.

Our inside-out simulations are complementary to a number of works that
focused on radiative feedback from an external field (Umemura \& Ikeuchi
1984; Rees 1986; Bond, Szalay \& Silk 1988; Efstathiou 1992; Thoul \&
Weinberg 1996; Kepner, Babul \& Spergel; Barkana \& Loeb 1999; Kitayama
\& Ikeuchi 2000; Susa \& Umemura 2000, 2004; Kitayama et al. 2000, 2001;
Shapiro, Iliev \& Raga 2003, 2004).  While external radiation can easily
ionize and blow away the outer envelope of gas clouds, photo-evaporation
proceeds less effectively in the densest central part.  Shapiro et
al. (2004) indeed show that evaporation of an initially hydrostatic
mini-halo ($\sim 10^6 \msun$) takes about 100 million years when
irradiated by a luminous source at a distance of $\sim 1$ Mpc. At the
very center of the halo, self-shielding and dynamical infall of the gas
can prevent complete evaporation and promote star formation (Kitayama et
al. 2001; Susa \& Umemura 2004).  The subsequent evolution of the gas
cloud will then be largely regulated by the feedback from the central
star.  As will be shown in the present paper, ionization of the ambient
gas by an internal source is more rapid than that by the external
radiation because of the greater incident radiation flux and the
presence of the gas density gradient.

The final state of the ionized gas near the central region is of
particular interest in the study of the first supernova explosions
(e.g. Bromm, Yoshida \& Hernquist 2003; Wada \& Venkatesan 2003).  The
subsequent evolution of the ionized gas has also an important
implication for the formation of proto-galaxies (Oh \& Haiman 2003).
For the gas clouds hosted by mini-halos with mass $\sim 10^6 \msun$,
one may naively expect that the gas is completely photoionized and
eventually photoevaporated, because the virial temperature of the
system is much lower than the characteristic temperature of
photoionized gas.  On the other hand, larger dark matter halos generate
a deeper gravitational potential well, and thus it can effectively trap
the hot, ionized gas within a small radius, rather than letting it move
outward by pressure.  The actual situation is likely to be far more
complicated, depending upon the luminosity and the spectrum of the
radiation source, and also on the initial gas density profile. It is
clearly important to make reliable predictions, under various physical
conditions, as to how the gas is re-distributed by radiation from the
central star.

Throughout the present paper, we work with a $\Lambda$-dominated CDM
cosmology with the matter density $\Omega_{\rm M}=0.3$, the cosmological
constant $\Omega_{\Lambda}=0.7$, the Hubble constant $h =0.7$, and the
baryon density $\Omega_{\rm B}=0.05$.

\section{The simulations}

\subsection{Numerical Scheme}

We study the evolution of \HII regions around a primordial star using
the radiation-hydrodynamics code of Kitayama et al. (2001).  The code
employs the second-order Lagrangian finite-difference scheme in
spherically symmetric geometry (Bowers \& Wilson 1991; see also Thoul \&
Weinberg 1995).  It treats self-consistently gravitational force,
hydrodynamics, non-equilibrium chemistry of primordial gas, and the
radiative transfer of UV photons. In the present paper, we further
incorporate the radiation force. We also adopt an artificial viscosity
formulation of Caramana, Shashkov \& Whalen (1998), designed to
distinguish between shock-wave and uniform compression using an
advection limiter.

The basic equations are given by
%%%%%%%%%%%%%%%%%%%%%
\begin{eqnarray}
\frac{dm}{dr} &=& 4 \pi r^2 \rho, \\
\label{eq:mom}
\frac{d^2r}{dt^2} &=& -4 \pi r^2 \frac{dp}{dm} -
\frac{G M_{\rm tot}(<r)}{r^2} + f_{\rm rad}, \\ 
\label{eq:eng}
\frac{du}{dt} &=& \frac{p}{\rho^2}\frac{d\rho}{dt} + \frac{{\cal H 
    - L}}{\rho}, \label{eq:energy}\\ 
p&=& \frac{2}{3} \rho u 
\end{eqnarray}
%%%%%%%%%%%%%%%%%%%%%
where $r$, $m$, $\rho$, $p$, $u$, and $M_{\rm tot}(<r)$ are the radius,
mass, density, pressure, internal energy per unit mass, and the total
mass inside $r$, respectively.  The radiative heating and cooling rates
per unit volume, ${\cal H}$ and ${\cal L}$, and the radiation force per
unit mass, $f_{\rm rad}$, depend on the solutions of the radiative
transfer equation and the chemical reactions.  We therefore take the
following procedure at each timestep we advance with the momentum
equation (eq. [\ref{eq:mom}]).

First, the direction and frequency dependent radiative transfer is
worked out as described in detail in Appendix A. We solve both
absorption and emission of ionizing ($\geq 13.6$ eV) photons and take
account of self-shielding of H$_2$ against the Lyman-Werner (LW,
11.2--13.6 eV) band photons following Draine \& Bertoldi (1996).  This
yields the UV heating rate ${\cal H}$ and the radiation force $f_{\rm
rad}$, together with the coefficients for photoioniziation of H,
photodissociation of H$_2$ and H$_2^+$, and photodetachment of
H$^-$. The obtained coefficients are used in the chemical reaction
network mentioned below.

Secondly, non-equilibrium chemical reactions are solved by an implicit
scheme developed by Susa \& Kitayama (2000) for the species e, H, H$^+$,
H$^-$, H$_2$, and H$_2^+$.  \footnote{In the present paper, we consider
only the hydrogen component of the IGM. Including helium will raise the
temperature of photoionized gas and may affect the overall evolution of
\HII regions. This point will be discussed in \S \ref{sec:mcrit}.}
Unless otherwise stated, the reactions and the rates are taken from the
compilation of Galli \& Palla (1998).

Finally, the energy equation (eq. [\ref{eq:eng}]) is solved including UV
heating and radiative cooling due to collisional ionization, collisional
excitation, recombination, thermal bremsstrahlung, Compton scattering
with the cosmic microwave background (CMB) radiation, and
rotational-vibrational excitation of H$_2$. The atomic cooling rates are
taken from the compilation of Fukugita \& Kawasaki (1994) and molecular
cooling rates from Galli \& Palla (1998).

The above procedure is repeated iteratively until the abundance of each
species and the internal energy in each mesh converge within an accuracy
of 0.1\%. We have validated the accuracy of our code by carrying out a
suite of conventional tests.  The results of a ``Str\"omgren sphere''
test are presented in Appendix B.

\subsection{Central Source}

At the center of a gas cloud, we place a metal-free Population III
star. Its parameters are taken from Schaerer (2002) and listed in Table
\ref{tab-star}. We vary the stellar mass, denoted by $M_{\rm star}$,
over the range $25 \sim 500 \msun$ and approximate the spectrum by a
black-body with the effective temperature $T_{\rm eff}$. The luminosity
is normalized so that the number of ionizing photons emitted per second,
$\dot{N}_{\rm ion}$, matches the time averaged photon flux given in
Table 5 of Schaerer (2002).  For simplicity, we neglect its spectral
evolution. Unless stated otherwise, the simulations are performed during
the main sequence lifetime, $t_{\rm life}$, of the central star.

%%%%%%%%%%%%%%%%%%%%%%%%%%%%%%%
\begin{table}
\caption{Properties of metal-free stars (Schaerer 2002).}
\label{tab-star}
\begin{center}
\begin{tabular}{cccc}
\hline \\[-9pt] 
\hline \\[-6pt] 
$M_{\rm star}$ [$\msun$] & $t_{\rm life}$ [Myr] & $T_{\rm eff}$ [K]& 
$\dot{N}_{\rm ion}$ [s$^{-1}$]\\[4pt]\hline \\[-6pt]
500 & 1.90 & $1.07 \times 10^5$ & $6.80 \times 10^{50}$ \\
200 & 2.20 & $9.98 \times 10^4$ & $2.62 \times 10^{50}$\\
80 & 3.01  & $9.33 \times 10^4$ & $7.73 \times 10^{49}$\\
25 & 6.46 & $7.08 \times 10^4$ & $7.58 \times 10^{48}$\\[4pt] \hline 
\end{tabular} 
\end{center}
\end{table}
%%%%%%%%%%%%%%%%%%%%%%%%%%%%%%%

\subsection{Initial Conditions}

We assume that a star-forming gas cloud is embedded in a dark matter
halo with the NFW density profile (Navarro, Frenk \& White 1997):
\begin{equation}
\rho_{\rm DM}  \propto  
\frac{1}{x(1+x)^2},
\end{equation}
where $x=r/r_{\rm s}$ is the radius normalized by the scale radius
$r_{\rm s}$.  We follow Bullock et al. (2001) to determine $r_{\rm s}$
for a halo with total mass $M_{\rm halo}$ collapsing at redshift $z_{\rm
c}$, by extrapolating their formula to the lower halo masses and the
higher redshifts. The dark matter density is normalized so that the dark
matter mass within the virial radius is equal to $M_{\rm DM}= (1 -
\Omega_{\rm B}/\Omega_{\rm M}) M_{\rm halo}$. We assume that the dark
matter density profile remains unchanged, since the halo dynamical
timescale is much longer than the lifetime of the massive star. 

For the gas in the halo, we adopt a power-law density profile
\begin{equation}
\rho \propto r^{-w},
\end{equation}
and vary the index $w$ from 1.5 to 2.25. This power-law profile are
motivated by the results of recent three dimensional simulations of the
primordial gas cloud formation (Abel et al. 2002; Yoshida et al. 2003a)
which show that the gas density profile around the first star-forming
regions is well described with $w\sim 2$ over a wide range of radii.
The gas density is normalized so that the total gas mass within the halo
virial radius is equal to $M_{\rm gas}=(\Omega_{\rm B}/\Omega_{\rm M})
M_{\rm halo} - M_{\rm star}$.  We determine the radius of the inner-most gas
shell $r_{\rm in}$ such that
%%%%%%%%%%%%%%%%%%%%%%%%%%
\begin{equation}
\bar{n}_{\rm H}(< r_{\rm in}) = 10^6 {\rm cm}^{-3}
\label{eq-rin}
\end{equation} 
%%%%%%%%%%%%%%%%%%%%%%%%%%
where $\bar{n}_{\rm H}(< r)$ is the average hydrogen number density
within $r$.  This density is comparable to that of primordial molecular
cloud cores or fragments found in numerical simulations (Abel et
al. 2002; Bromm et al. 2002; Nakamura \& Umemura 2002), and gives
$r_{\rm in}$ sufficiently smaller than the characteristic size of the
simulated halos. Our primary interest hence lies in the propagation of
UV photons after they escape out of the dense molecular cloud.
\footnote{The infalling envelope may prevent the escape of UV photons
from dense proto-stellar clouds. Omukai \& Inutsuka (2001) showed that,
in spherically symmetric cases, \HII regions around a massive star
cannot expand under significant mass accretion.  Mass accretion onto a
dense proto-stellar cloud, however, could occur along an aspherical disk
due to its angular momentum. For disk-like geometries, escape of the UV
photons from the proximity of the massive star will be greatly enhanced
than in the spherically symmetric case.  In the present paper, we focus
on such cases that the UV photons could successfully escape out of a
proto-stellar disk and propagate through the surrounding halo.}  If a
gas shell falls below $0.5 r_{\rm in}$, it is moved to the center and
ignored in the rest of the simulation, except in the calculation of the
gravitational force.  In order to trace the propagation of an I-front
for a sufficiently long time, the outer boundary is taken at $1 \sim 3$
times the virial radius depending on the run. The gas shells are
initially spaced such that the shell mass increases by a constant ratio,
typically $\sim 1\%$, between the adjacent shells.  We have checked that
a total of 300 radial bins from the center to the outermost shell is
sufficient to produce converged results. Since the star-forming clouds
are likely to be in the course of collapse, we assign the initial
velocity
%%%%%%%%%%%%%%%%%%%%%%%%%%
\begin{equation}
v_{\rm init}(r) = - \sqrt{\frac{2 G M_{\rm tot}(<r)}{r} }. 
\end{equation}
%%%%%%%%%%%%%%%%%%%%%%%%%%

\epsscale{1.07}
%%%%%%%%%%%%%%%%%%%%%%%%%%%%%%%%%%%%%%%%%%%
\begin{figure*}
\plotone{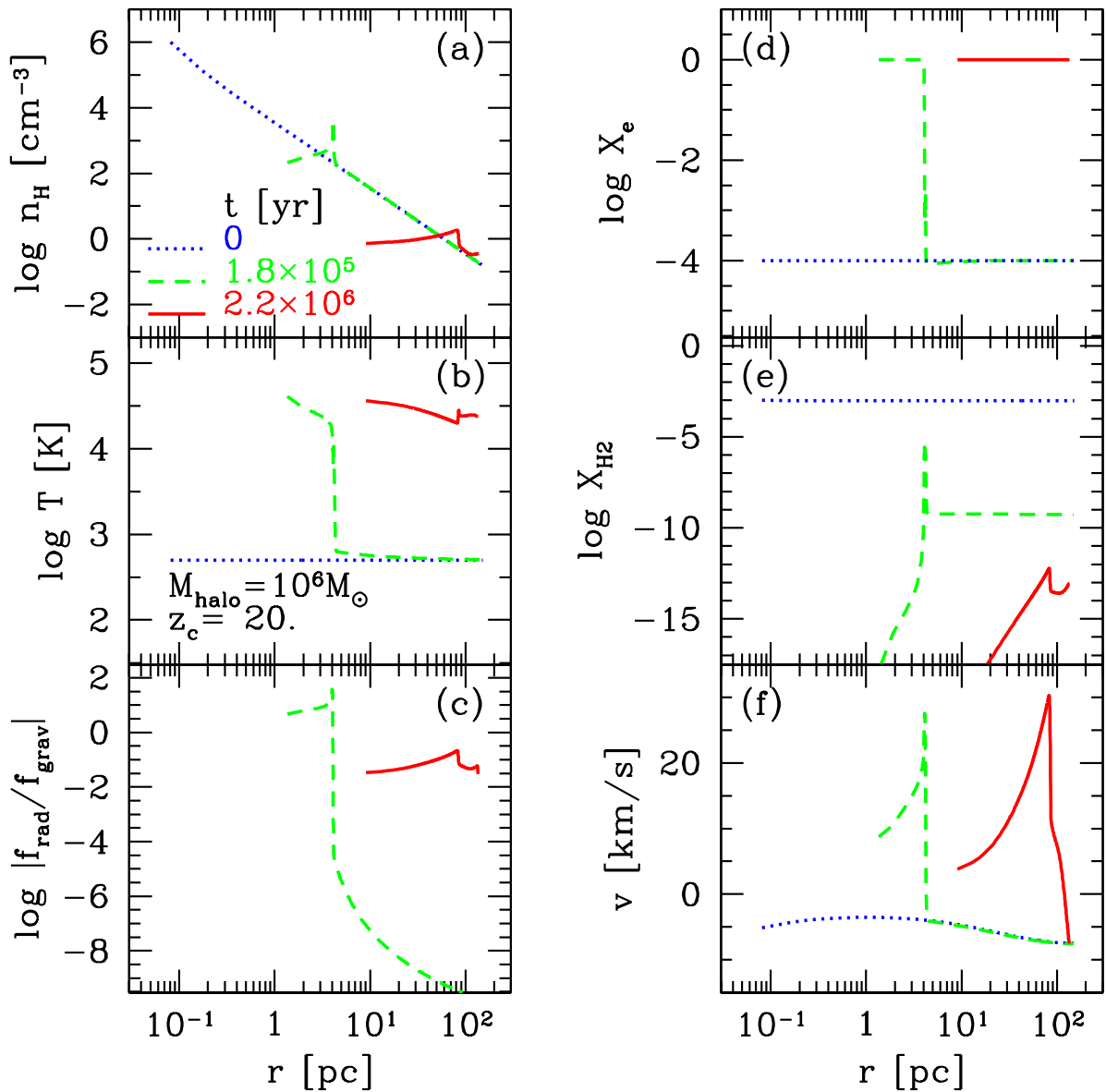} \caption{Structure of an \HII region around a massive
star with $M_{\rm star}=200 \msun$ inside a halo with $M_{\rm
halo}=10^6$ M$_\odot$ and $w=2.0$ at $z_{\rm c}=20$.  Radial profiles
are shown at $t=0$ (dotted lines), $1.8 \times 10^5$ yr (dashed lines),
and $2.2 \times 10^6$ yr (solid lines) for (a) hydrogen density, (b)
temperature, (c) ratio of radiation force to gravitational force, (d)
electron fraction, (e) H$_2$ fraction, and (f) radial velocity.
\label{fig:prof1}}
\end{figure*}
%%%%%%%%%%%%%%%%%%%%%%%%%%%%%%%%%%%%%%%%%%%%%%
%%%%%%%%%%%%%%%%%%%%%%%%%%%%%%%%%%%%%%%%%%%
\begin{figure*}
\plotone{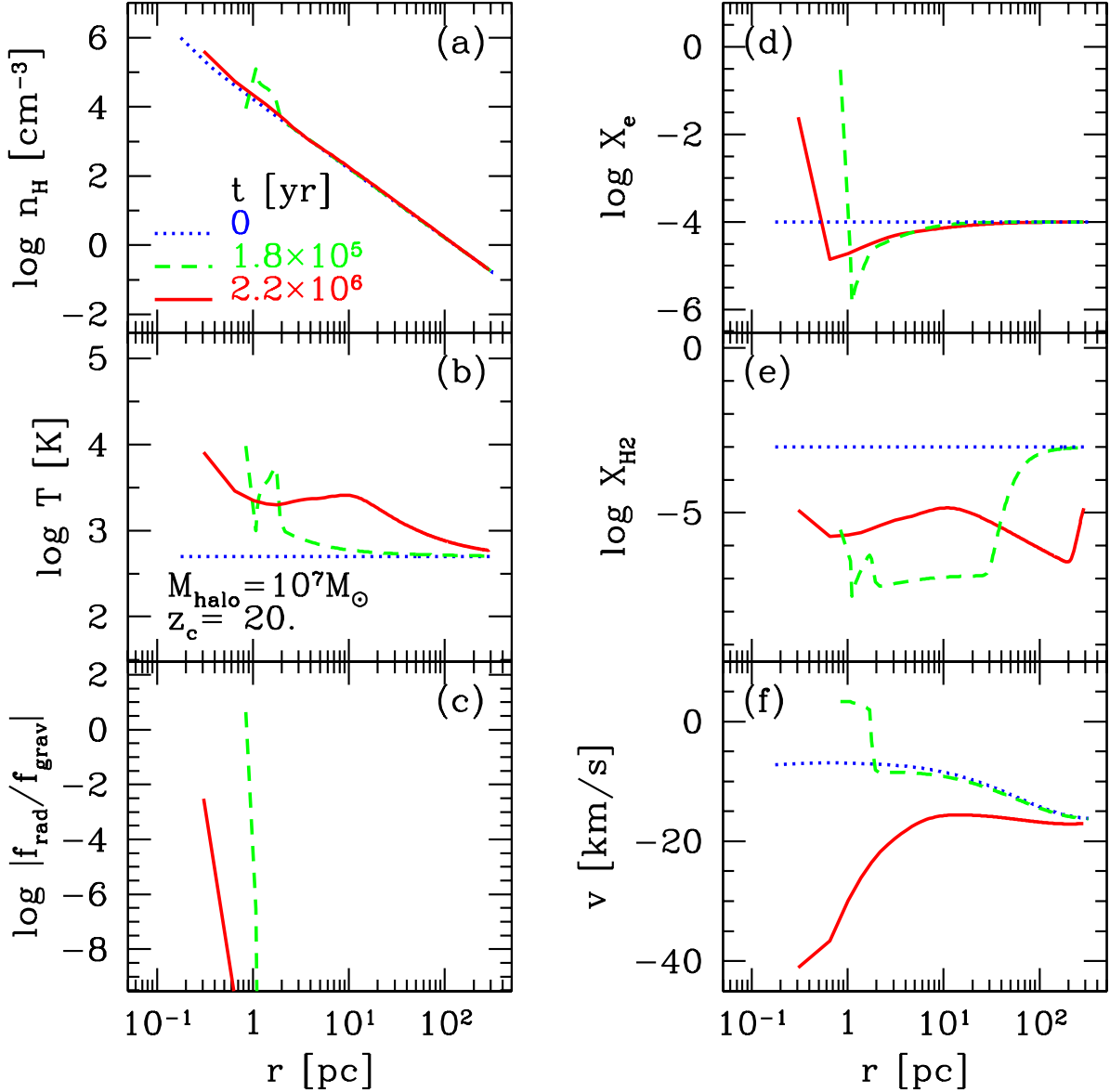}
\caption{Same as Figure \ref{fig:prof1} except that 
the halo mass is $M_{\rm halo}=10^7$ M$_\odot$. 
\label{fig:prof2}}
\end{figure*}
%%%%%%%%%%%%%%%%%%%%%%%%%%%%%%%%%%%%%%%%%%%%%%
%%%%%%%%%%%%%%%%%%%%%%%%%%%%%%%%%%%%%%%%%%%
\begin{figure*}
\plotone{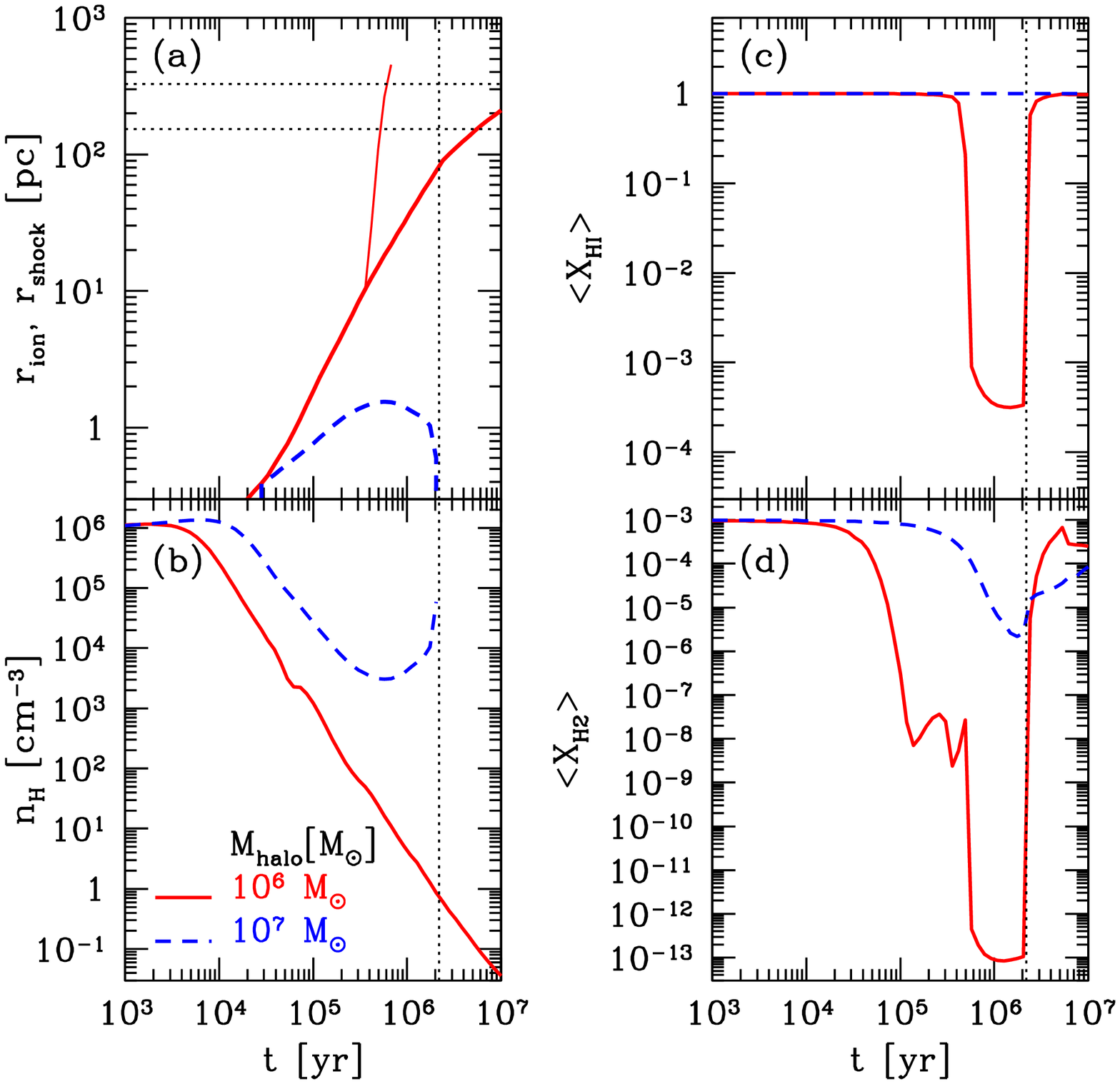} \caption{Evolution of (a) shock front radius (thick
lines) and I-front radius (thin lines, whenever separated from the
shock), (b) central hydrogen density, (c) mean HI fraction within
$r_{\rm vir}$, and (d) mean H$_2$ fraction within $r_{\rm vir}$, for
halos with $M_{\rm halo}=10^6$ M$_\odot$ (solid lines) and $10^7$
M$_\odot$ (dashed lines) shown in Figures \ref{fig:prof1} and
\ref{fig:prof2}, respectively.  Horizontal dotted lines denote virial
radii of halos with $M_{\rm halo}=10^6$ M$_\odot$ and $10^7$ M$_\odot$,
while the vertical one indicates the lifetime of the central star with $M_{\rm
star}=200 \msun$.  \label{fig:evol}}
\end{figure*}
%%%%%%%%%%%%%%%%%%%%%%%%%%%%%%%%%%%%%%%%%%%%%%

We assume that the gas is initially isothermal with temperature $T_{\rm
init}$, given by the {\it minimum} of the virial temperature $T_{\rm
vir}$ and a reference temperature $T_{\rm min}$. We take $T_{\rm min} =
500$ K, corresponding to the gas which has cooled by molecular hydrogen
cooling. The final results are found to be insensitive to the choice of
$T_{\rm min}$, because, as long as $T_{\rm min} \ll 10^4$K, the gas is
initially almost neutral and the opacity to ionizing photons is very
large.
%For $T_{\rm min} > 10^4$ K (relevant for $T_{\rm vir} >
%10^4$K), the gas evolution is little affected by photoionization
%regardless of the initial temperature. 
The initial abundances are taken to be consistent with the above choice
of the initial temperature based on the hydrodynamical simulations of
Kitayama et al. (2001); $X_{\rm e}=10^{-4}$, $X_{{\rm H}^-} =10^{-10}$,
$X_{\rm H2}=10^{-3}$, and $X_{\rm{H2}^+}=10^{-12}$, where $X_{\rm
i}\equiv n_{\rm i}/n_{\rm H}$ is the fraction of the species i with
respect to the total number of hydrogen atoms. 

\section{Results}

\subsection{I-front propagation} 

We first describe in detail the results of our fiducial runs with
$M_{\rm star}=200 \msun$, $w=2.0$, and $z_{\rm c}=20$.  Simulations with other
sets of parameters are also carried out and the results are presented in
due course.

Figure \ref{fig:prof1} shows the structure of the \HII region at $t=0$
(initial), $1.8\times 10^5$, and $2.2\times 10^6$ yr for a low mass case
with $M_{\rm halo}=10^6 \msun$.  The evolution in the early stage is
characterized by the propagation of a weak D-type front into the
surroundings; shock precedes the I-front. Because of the steep density
gradient, the I-front eventually overtakes the shock and changes into
R-type. The transition takes place at $t\sim 3\times 10^5$ yr (see also
Figure \ref{fig:evol} and discussions below).  In this so-called
``champagne'' phase, the low-density envelope is promptly ionized
(Welter 1980) and the gas temperature rises to a few times $10^4$~K.  As
the shock propagates at the speed $\sim 30$ km s$^{-1}$, it reaches
$\sim 70$ pc within the lifetime of the massive star, $t_{\rm life}$=2.2
Myr. 

The spatial resolution of the present simulation is sufficient to
resolve the cooling layer behind the shocks; e.g., the I-front at
$t=1.8\times 10^5$ yr in Figure \ref{fig:prof1} is resolved with more
than 10 shells.  We have also performed the runs with halving and
doubling the gas shell numbers and confirmed that the results are almost
identical.

We notice that there appears a thin shell of H$_2$ just in front of the
\HII region (Fig.~\ref{fig:prof1}e) as pointed out by Ricotti et
al. (2001). This is due to a positive feedback on the H$_2$ formation by
an enhanced electron fraction at the temperature below $10^4$K.  As the
\HII region expands, the temperature in the shell exceeds $10^4$K and
H$_2$ molecules are dissociated by collisions with ions.  There still
appears a new H$_2$ shell in front of the I-front, but in a different
position (both Eulerian and Lagrangian) from the previous one.  The
H$_2$ shell is thus likely to be short-lived as a result of I-front
propagation.  There may also be a case that the H$_2$ formation is
promoted after the central star fades away and the gas cools to
temperatures below $10^4$K.  Such a possibility will be explored later
in this subsection.

Interestingly, the radiation force dominates over the gravitational
force inside the \HII region when the I-front radius is smaller than
$\sim 30$ pc (Fig.~\ref{fig:prof1}c).  This value is a factor of
$\sim 4$ smaller than that indicated by Haehnelt (1995), and easily
accounted for by the fact that there is about 3--5 times more mass
contributed by dark matter than baryons near the center in our
simulations.

For a higher mass halo with $10^7 \msun$ illustrated in Figure
\ref{fig:prof2}, the outward gas motion near the center is eventually
reverted to an inflow.  Correspondingly, the shock radius starts {\it
decreasing}.  This reversion is explained by a combination of the deep
gravitational potential well near the halo center and the infall of the
envelope gas.  Notice that the ionization fraction decreases sharply at
$r \sim 1$ pc and $t=1.8\times 10^5$ yr.  The infalling material piles on
the shock front, enhancing recombination (Fig. \ref{fig:prof2}d).  This
is however just in a transition phase; the piled gas rapidly falls back
to the center. Thereafter, the \HII region is kept trapped within a few
parsec radius because the gas density remains high ($\ga 10^{4} {\rm
cm}^{-3}$) and recombination balances photoionization.  We have carried
out a simulation for the same set of parameters but assuming that the
gas is initially static. Even in this case, the D-type front maintains
over $t_{\rm life}$=2.2 Myr and reaches only $\sim 40$~pc, still well
inside the virial radius.

Figure \ref{fig:evol} further compares the time evolution in the two
cases presented in Figures \ref{fig:prof1} and \ref{fig:prof2}.  For
definiteness, we take the shock front radius at the peak of gas density
and the I-front radius at the position with $X_{\rm HI}=0.1$.  The mass
weighted mean fractions of HI and H$_2$ are computed within the virial
radius $r_{\rm vir}$.  For $M_{\rm halo}=10^6\msun$, the I-front
detaches from the shock at $t\sim 3\times 10^5$ yr and then propagates
rapidly, exceeding the halo virial radius.  The halo gas is almost
completely ionized in the first few hundred thousand years and then a
large fraction of ionizing and the LW photons can escape from the halo.
For $M_{\rm halo}=10^7 \msun $, on the other hand, the I-front {\it
stalls} at $t\sim 3\times 10^5$ yr and then moves back inward.  A large
fraction of the gas is neutral but the H$_2$ fraction drops by three
orders of magnitude from the initial value because the LW photons can
penetrate further than ionizing ones.

The evolution after the lifetime of the central star ($t > 2.2$ Myr) is
also plotted in Figure \ref{fig:evol}. In the $10^6 \msun$ case, if the
central star simply fades away without triggering a supernova explosion
(as is assumed in our calculations), the gas is undisturbed and quickly
recombine (Fig.~\ref{fig:evol}c) and reform molecules
(Fig.~\ref{fig:evol}d).  The large ionization fraction within the \HII
region can accelerate production of hydrogen molecules through H$^{-}$
formation. The shock front continues expanding until it dissipates
kinetic energy.

Alternatively, if the central star turns into a supernova, it will
generate a blastwave, which runs through the ambient gas. A quantity
that is of particular importance in such a case is the final density of
the evacuated gas near the center (see \S \ref{sec:imp} for more
details). It is in fact primarily determined by the host halo mass. In
the low-mass case ($M_{\rm halo}=10^6\msun$), the central gas density
drops to $\sim 1$ cm$^{-3}$ at $t=2.2$ Myr. In the high-mass case
($M_{\rm halo}=10^7\msun$), the central density remains much higher at
$\ga 10^4 {\rm cm}^{-3}$.  The threshold between the two cases closely
follows that of the escape fraction of ionizing photons described in \S
\ref{sec:mcrit}.

\epsscale{1.6}
%%%%%%%%%%%%%%%%%%%%%%%%%%%%%%%%%%%%%%%%%%%
\begin{figure}
\hspace*{-1.4cm} \plotone{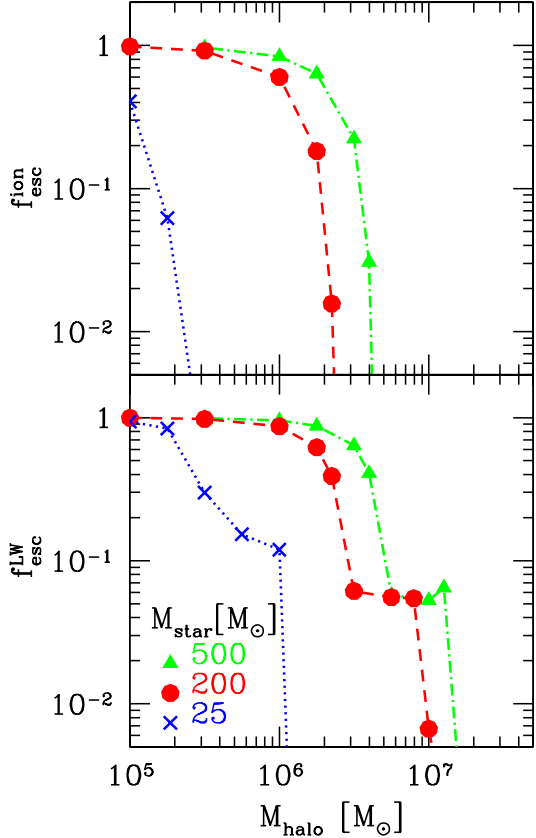} 
\caption{ Escape fractions of ionizing photons ($\geq
13.6$ eV, top panel) and the LW photons (11.2--13.6 eV, bottom panel)
versus $M_{\rm halo}$.  Symbols denote stellar masses of $M_{\rm
star}=500$ M$_\odot$ (triangles), 200 M$_\odot$ (circles) and 25
M$_\odot$ (crosses).  A plateau and a small dip at $f_{\rm esc}^{\rm LW}
\sim 0.1$ are due to a weak positive feedback on H$_2$ formation by an
enhanced electron fraction.  \label{fig:fesc_ms}}
\end{figure}
%%%%%%%%%%%%%%%%%%%%%%%%%%%%%%%%%%%%%%%%%%%%%%
%%%%%%%%%%%%%%%%%%%%%%%%%%%%%%%%%%%%%%%%%%%
\begin{figure}
\hspace*{-1.4cm} \plotone{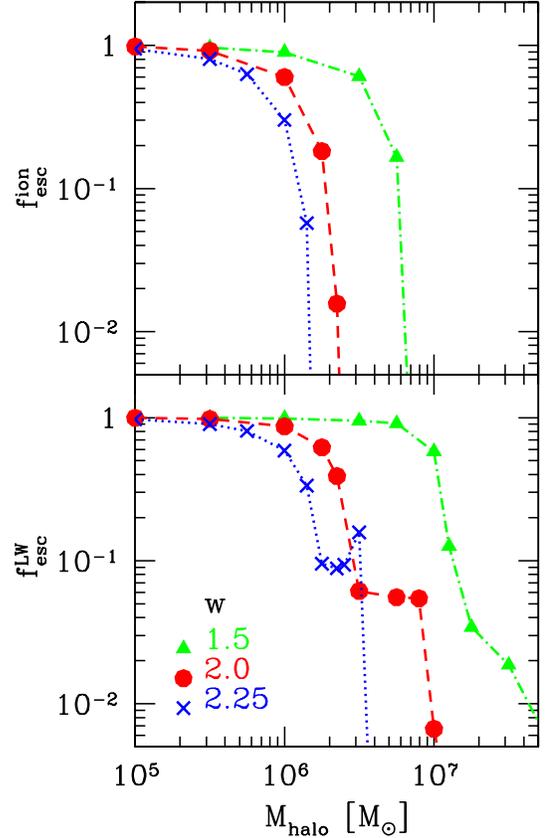} 
\caption{Same as Figure \ref{fig:fesc_ms} except that
symbols denote gas density slopes of $w=1.5$ (triangles) 2.0 (circles)
and 2.25 (crosses).  \label{fig:fesc_w}}
\end{figure}
%%%%%%%%%%%%%%%%%%%%%%%%%%%%%%%%%%%%%%%%%%%%%%
%%%%%%%%%%%%%%%%%%%%%%%%%%%%%%%%%%%%%%%%%%%
\begin{figure}
\hspace*{-1.4cm} \plotone{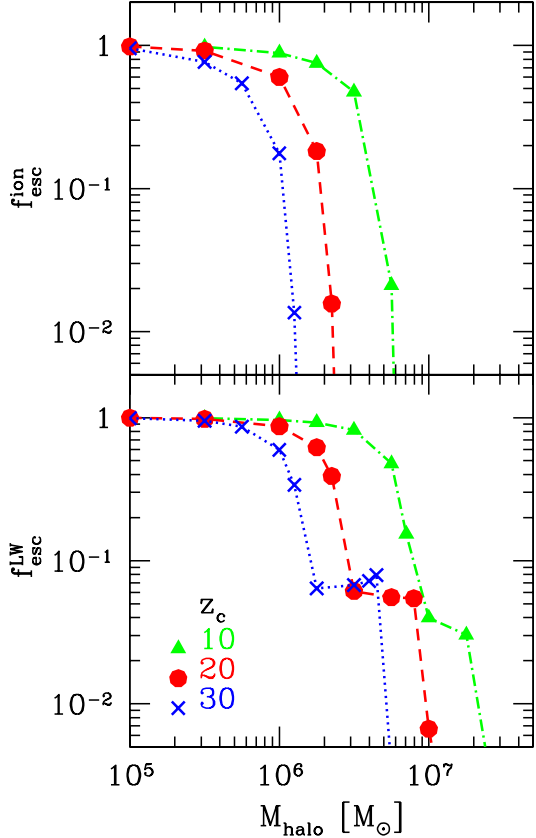} 
\caption{Same as Figure \ref{fig:fesc_ms} except that
symbols denote collapse redshifts of $z_{\rm c}=10$ (triangles), 20
(circles) and 30 (crosses).  \label{fig:fesc_zc}}
\end{figure}
%%%%%%%%%%%%%%%%%%%%%%%%%%%%%%%%%%%%%%%%%%%%%%

\subsection{Photon escape fraction}

A useful output of our simulations is the escape fractions of ionizing
photons and the LW photons from a halo. The former is of particular
significance in terms of early reionization of the Universe (e.g.,
Yoshida et al. 2003b; Sokasian et al. 2004; Whalen et al. 2004), while
the latter quantifies the efficiency of the so-called negative feedback
on subsequent star formation (Haiman, Rees \& Loeb 1997; Omukai \& Nishi
1999).

In Figures \ref{fig:fesc_ms}--\ref{fig:fesc_zc}, we plot the escape
fractions, averaged over the lifetime of the central star, as a function
of halo mass for various choices of $M_{\rm star}$, $w$, and $z_{\rm
c}$.  In the case of $M_{\rm star}=25 \msun$ (Figure \ref{fig:fesc_ms}),
for instance, the escape fraction of ionizing photons $f_{\rm esc}^{\rm
ion}$ is below 0.5 over the range of $M_{\rm halo}$ we consider.  For
$M_{\rm star}=200 \msun$, on the other hand, $f_{\rm esc}^{\rm ion}$ is
close to unity at $M_{\rm halo}< 10^6 \msun$ but drops sharply at
$M_{\rm halo}\sim 2\times 10^6 \msun$.  There appears to be a critical
mass of the host halo above which $f_{\rm esc}^{\rm ion}$ is essentially
zero.  For the escape fraction of the LW photons $f_{\rm esc}^{\rm LW}$,
similar features are found with a slightly larger critical mass, because
the LW photons are harder to shield than ionizing photons.  A plateau
and a small dip at $f_{\rm esc}^{\rm LW} \sim 0.1$ are real; a minor
leakage of ionizing photons causes a weak positive feedback on H$_2$
formation by enhancing electron fraction (Haiman, Rees \& Loeb 1996).

Figure \ref{fig:fesc_w} indicates that a steeper density profile leads
to smaller escape fractions.  This is because, for given gas mass in a
halo, a steeper profile results in higher density near the center. The
I-front propagation is then prevented at early stages due to the higher
recombination rate.  Similarly, an earlier collapse epoch implies a
denser, compact halo, from which escape fractions are smaller (Figure
\ref{fig:fesc_zc}).

Strictly speaking, the self-shielding function of the LW photons we
adopt (Drain \& Bertoldi 1996) is derived for the stationary gas.  In
the presence of supersonic flows, the photons may doppler-shift out of
or into particular lines in the LW band. Detailed account for these
effects requires precise treatment of the line transfer incorporating
velocity and temperature structures of the gas, which is beyond the scope
of the present paper.  Our results on $f_{\rm esc}^{\rm LW}$ should
therefore be regarded as a minimal estimation for the true value.

\epsscale{1.0}
%%%%%%%%%%%%%%%%%%%%%%%%%%%%%%%%%%%%%%%%%%%
\begin{figure}
\plotone{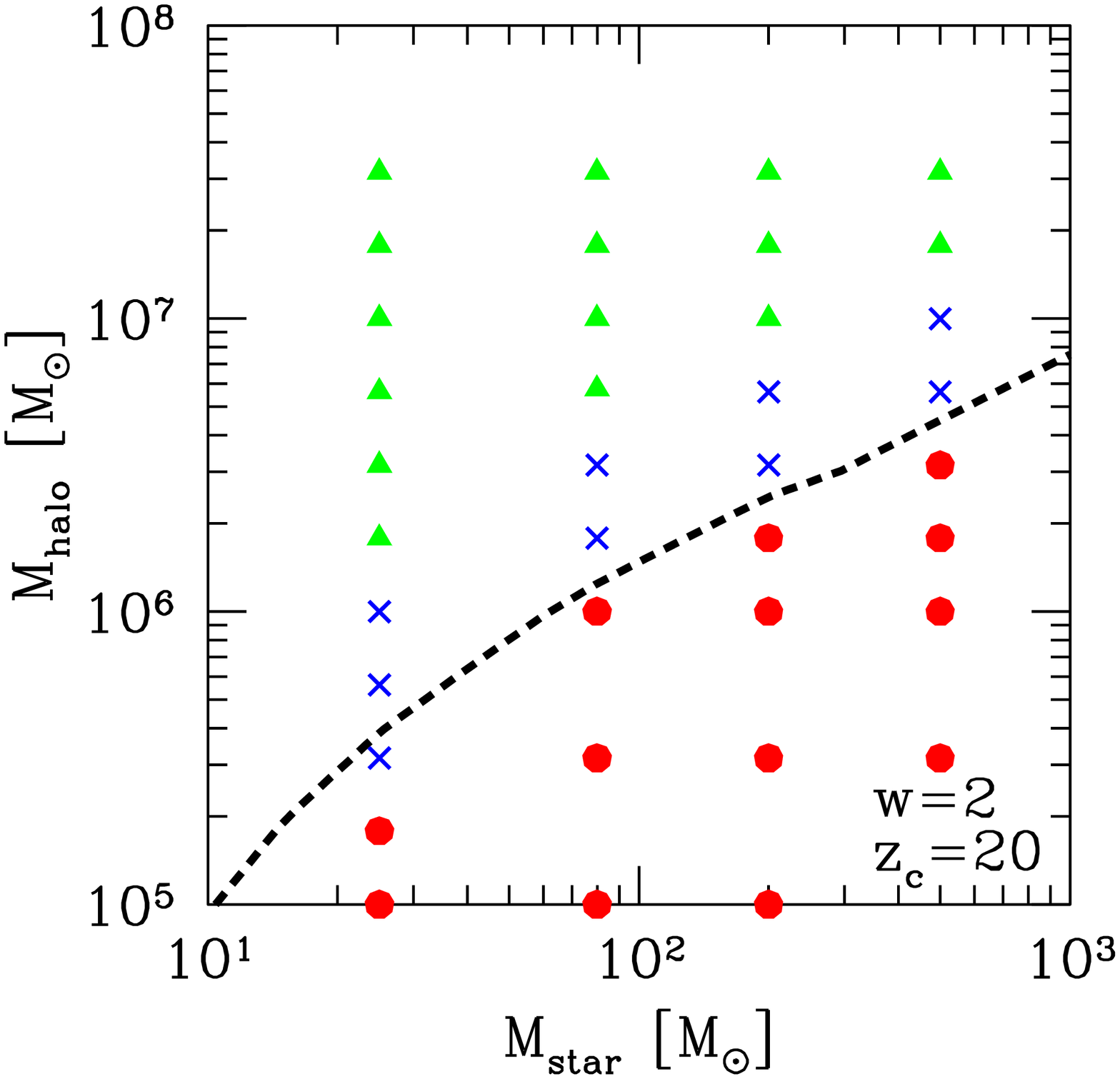} \plotone{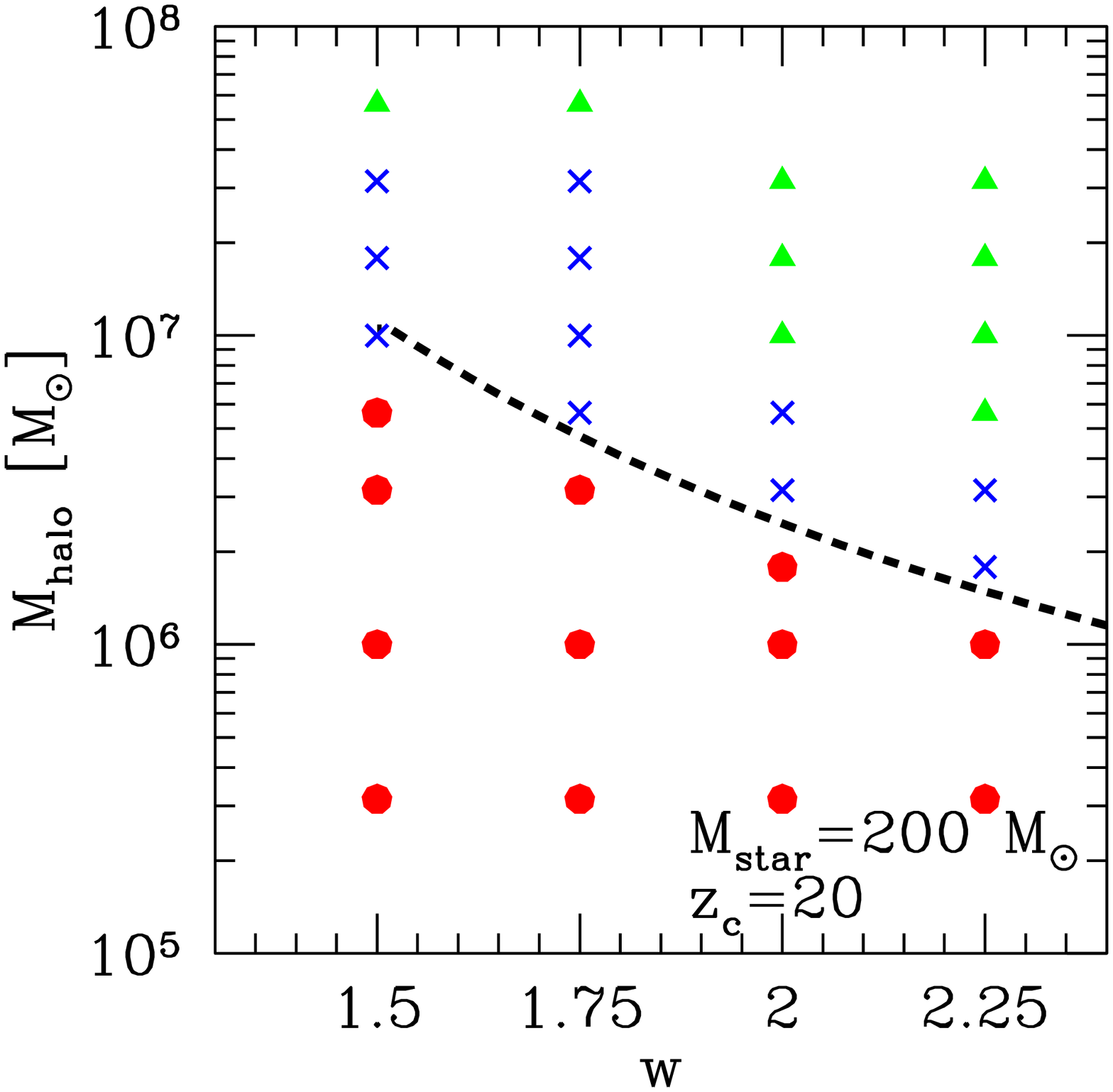}
\plotone{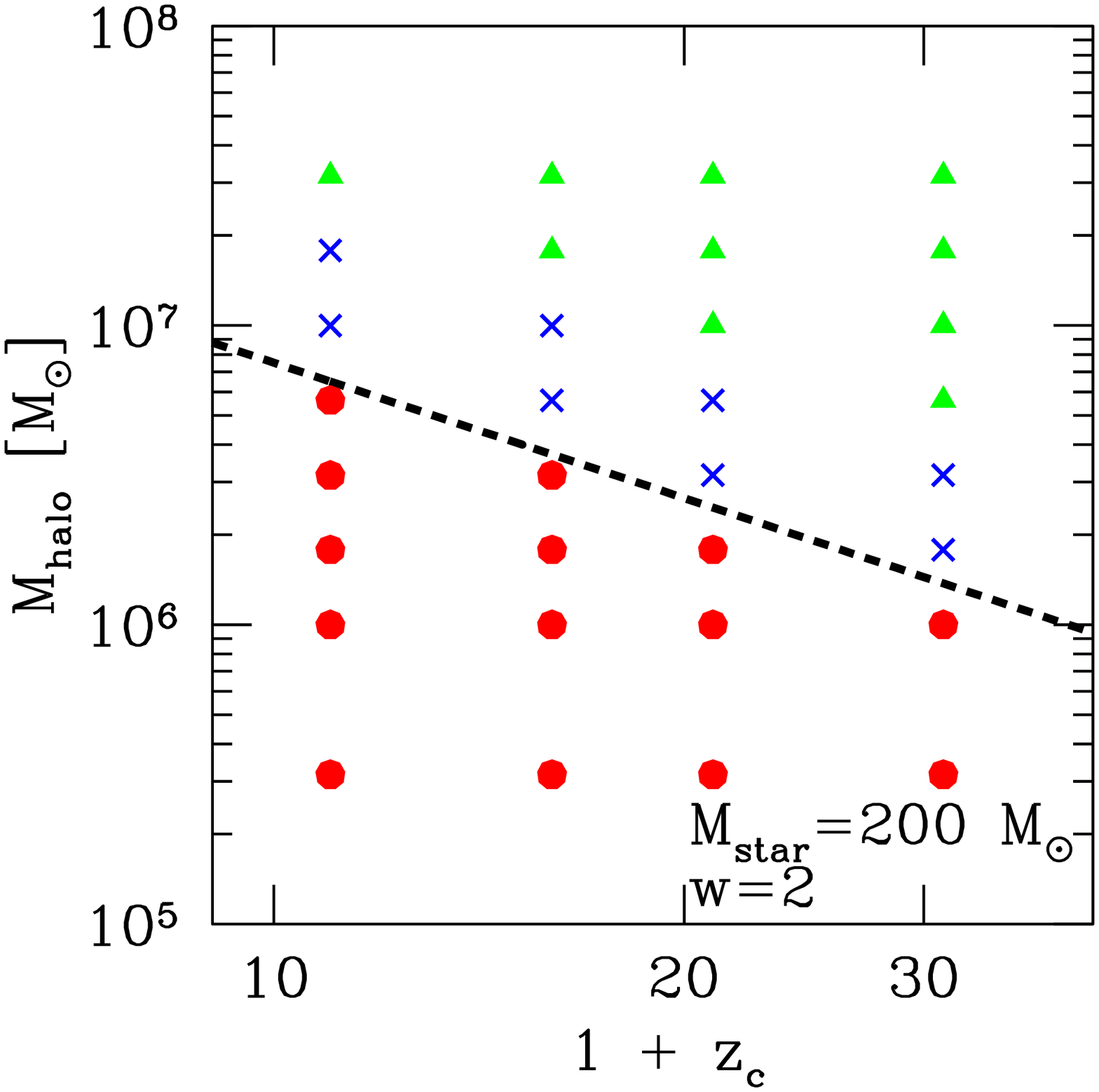}
\caption{Critical masses of escape fractions as a function of $M_{\rm
star}$, $w$, and $z_{\rm c}$.  Circles denote $f_{\rm esc}^{\rm ion} >
1\%$ and $f_{\rm esc}^{\rm LW} > 1\%$, crosses $f_{\rm esc}^{\rm ion} <
1\%$ and $f_{\rm esc}^{\rm LW} > 1\%$, and triangles $f_{\rm esc}^{\rm
ion} < 1\%$ and $f_{\rm esc}^{\rm LW} < 1\%$.  Dashed lines indicate the
analytic estimation for the critical mass scale of complete ionization
(eq.  [\ref{eq-mcrit}]).  \label{fig:msm}}
\end{figure}
%%%%%%%%%%%%%%%%%%%%%%%%%%%%%%%%%%%%%%%%%%%%%%

\subsection{Critical mass for photon escape}
\label{sec:mcrit}

Figure \ref{fig:msm} summarizes the degree of radiative feedback and
photon escape as a function of $M_{\rm star}$, $w$, and $z_{\rm c}$. Low
mass halos are ionized promptly after the onset of the central star,
resulting in high escape fractions of both ionizing and the LW
photons. On the other hand, the \HII regions are heavily confined and
the escape fractions are essentially zero in high mass halos. For
intermediate halos, a significant fraction of the LW photons can escape
from the halos while the ionizing photons are trapped. Figure
\ref{fig:msm} further indicates that the threshold mass scale for the
escape of ionizing photons, as a function of $M_{\rm star}$, $w$, and
$z_{\rm c}$, is well reproduced by the following analytic estimation.

Over the range of gas density gradient we consider ($1.5 < w < 2.25$),
the shock velocity is a weak function of $w$ and given roughly by $25 
\sim 35$ km s$^{-1}$ (see also Shu et al. 2002). For the low-mass halos
considered in this paper, the gravitational infall velocity is $5 \sim 10$
km s$^{-1}$.  Within the lifetime of the central star $t_{\rm life}$,
the D-type front can then reach the distance
%%%%%%%%%%%%%%%%%%%%%%%%%%%%%%%%%%
\begin{equation}
 l_{\rm D} = 20 \left(\frac{t_{\rm life}}{10^6 \mbox{yr}} \right) 
\left(\frac{v_{\rm D}}{20 \mbox{~km s$^{-1}$}} \right) \mbox{pc}, 
\label{eq-ld}
\end{equation}
%%%%%%%%%%%%%%%%%%%%%%%%%%%%%%%%%%
where $v_{\rm D}$ is the expansion velocity of a D-type front.  This is
in general much smaller than the virial radius of a halo in
consideration:
%%%%%%%%%%%%%%%%%%%%%%%%%%%%%%%%%%
\begin{equation}
 r_{\rm vir} = 160 
\left(\frac{M_{\rm halo}}{10^6 \mbox{M}_\odot} \right) ^{1/3} 
 \left(\frac{1+z_{\rm c}}{20} \right) ^{-1} \mbox{pc},  
\end{equation}
%%%%%%%%%%%%%%%%%%%%%%%%%%%%%%%%%%
applicable to current cosmology at $z \ga 10$.  It is therefore
necessary for the I-front to become R-type in order to ionize the whole
halo.

As shown in Figure \ref{fig:prof1}, the gas density inside the shock front
$r_{\rm s}$ is nearly constant and approximated by the average density
within $r_{\rm s}$:
%%%%%%%%%%%%%%%%%%%%%%%%%%%%%%%%%%
%\begin{equation}
% n_{\rm s} = \frac{3}{3-w} n_{\rm i}(r_{\rm s}) = 4.8 \times 10^{-5} (1+z)^3
%\left(\frac{r_{\rm s}}{r_{\rm vir}} \right) ^{-w} 
%\mbox{cm$^{-3}$},   
%\end{equation}
\begin{equation}
n_{\rm s} = \frac{3}{3-w} n_{\rm i}(r_{\rm s}) = 0.39 
\left(\frac{1+z_{\rm c}}{20}\right)^3
\left(\frac{r_{\rm s}}{r_{\rm vir}} \right) ^{-w} 
\mbox{cm$^{-3}$},   
\end{equation}
%%%%%%%%%%%%%%%%%%%%%%%%%%%%%%%%%%
where $n_{\rm i}(r_{\rm s})$ is the initial gas density at $r_{\rm s}$,
normalized by the condition that the total gas mass within $r_{\rm vir}$
is equal to $M_{\rm halo} \Omega_{\rm B}/\Omega_{\rm M}$.  One can
then define the Str\"omgren radius (Str\"omgren 1939) corresponding to
$n_{\rm s}$ as
%%%%%%%%%%%%%%%%%%%%%%%%%%%%%%%%%%
\begin{equation}
 r_{\rm St} = 150 \left(\frac{\dot{N}_{\rm ion}}{10^{50} \mbox{s}^{-1}} \right) ^{1/3} 
 \left(\frac{n_{\rm s}}{\mbox{cm}^{-3}} \right) ^{-2/3} \mbox{pc}. 
\end{equation}
%%%%%%%%%%%%%%%%%%%%%%%%%%%%%%%%%%
For $w >1.5$, $r_{\rm s}$ depends on $n_{\rm s}$ more weakly ($\propto
n_{\rm s}^{-1/w}$) than $r_{\rm St}$ ($\propto n_{\rm s}^{-2/3}$).  As
far as $r_{\rm St} < r_{\rm s}$, as is the case in the initial stage of
expansion, the I-front is of D-type. As the shock propagates and $n_{\rm
s}$ decreases, $r_{\rm St}$ eventually overtakes $r_{\rm s}$ and the
I-front changes into R-type.

Denoting the radius at which $r_{\rm St}$ is equal to $r_{\rm s}$ by
$r_{\rm eq}$, the necessary condition for the transition into the R-type
front within the lifetime of the central star is $r_{\rm eq} < l_{\rm D}
$, namely,
%%%%%%%%%%%%%%%%%%%%%%%%%%%%%%%%%%
\begin{equation}
M_{\rm halo} < M_{\rm crit}^{\rm ion},
\end{equation}
%%%%%%%%%%%%%%%%%%%%%%%%%%%%%%%%%%
where 
%%%%%%%%%%%%%%%%%%%%%%%%%%%%%%%%%%
\begin{eqnarray}
M_{\rm crit}^{\rm ion} &=& 5.0 \times 10^6
\sbkt{\frac{l_{\rm D}}{270 {\rm ~pc}}}^{3- \frac{9}{2w}}
\sbkt{\frac{\dot{N}_{\rm ion}}{10^{50} \mbox{s}^{-1}}}^{\frac{3}{2w}} 
\nonumber \\
&& \times \sbkt{\frac{1 + z_{\rm c}}{20}}^{3- \frac{9}{w}} \msun. 
\label{eq-mcrit}
\end{eqnarray}
%%%%%%%%%%%%%%%%%%%%%%%%%%%%%%%%%%
Figure \ref{fig:msm} shows the above critical mass for complete
ionization, adopting $v_{\rm D}=20$ km s$^{-1}$ in equation
(\ref{eq-ld}). In our fiducial case with $M_{\rm star} = 200 \msun$
($\dot{N}_{\rm ion}=2.6\times 10^{50}$ s$^{-1}$, $t_{\rm life}$=2.2
Myr), $w=2$, and $z_{\rm c} = 20$, it yields $M_{\rm crit}^{\rm ion} =
2.5 \times 10^6 \msun$. Indeed it agrees with our simulation results
over a wide range of parameters.

If one is to estimate the critical mass for the escape of the LW photons
separately, Figure \ref{fig:msm} indicates that $M_{\rm crit}^{\rm LW}
\sim 3 M_{\rm crit}^{\rm ion}$ gives a reasonable fit to the simulation
results for $w \simeq 2$.  In the case of $ 80 \msun \la M_{\rm star}
\la 500 \msun$, $ 10 \la z_{\rm c} \la 30$ and $w =2$, useful
approximations for the critical masses are
%%%%%%%%%%%%%%%%%%%%%%%%%%%%%%%%%%
\begin{equation}
 M_{\rm crit}^{\rm ion} \sim 2.5 \times 10^6  
\sbkt{\frac{M_{\rm star}}{200 \msun}}^{3/4}  
\sbkt{\frac{1 + z_{\rm c}}{20}}^{-3/2} \msun, 
\end{equation}
%%%%%%%%%%%%%%%%%%%%%%%%%%%%%%%%%%
%%%%%%%%%%%%%%%%%%%%%%%%%%%%%%%%%%
\begin{equation}
 M_{\rm crit}^{\rm LW} \sim 7.5 \times 10^6  
\sbkt{\frac{M_{\rm star}}{200 \msun}}^{3/4}  
\sbkt{\frac{1 + z_{\rm c}}{20}}^{-3/2} \msun, 
\end{equation}
%%%%%%%%%%%%%%%%%%%%%%%%%%%%%%%%%%
where we have used a power-law fit to the results of Schaerer (2002),
$\dot{N}_{\rm ion} \propto M_{\rm star}^{1.2}$ and $t_{\rm life} \propto
M_{\rm star}^{-0.2}$, over the range $ 80 \msun \la M_{\rm star} \la 500
\msun$.

If helium is included in the calculation, the gas temperature will be
higher by a factor of $\sim 2$ for $T_{\rm eff} \sim 10^5$ K, because of
an increase in the net heating rate (Abel \& Haehnelt 1999). This will
increase the shock velocity and consequently $l_{\rm D}$ in equation
(\ref{eq-ld}) by $\sim 40\%$.  The critical mass $M_{\rm crit}$ will
then become higher by $\sim 30\%$ for $w=2$.

\subsection{Effective clumping factor of halos}
\label{sec:clump}

%%%%%%%%%%%%%%%%%%%%%%%%%%%%%%%%%%%%%%%%%%%
\begin{figure}
\plotone{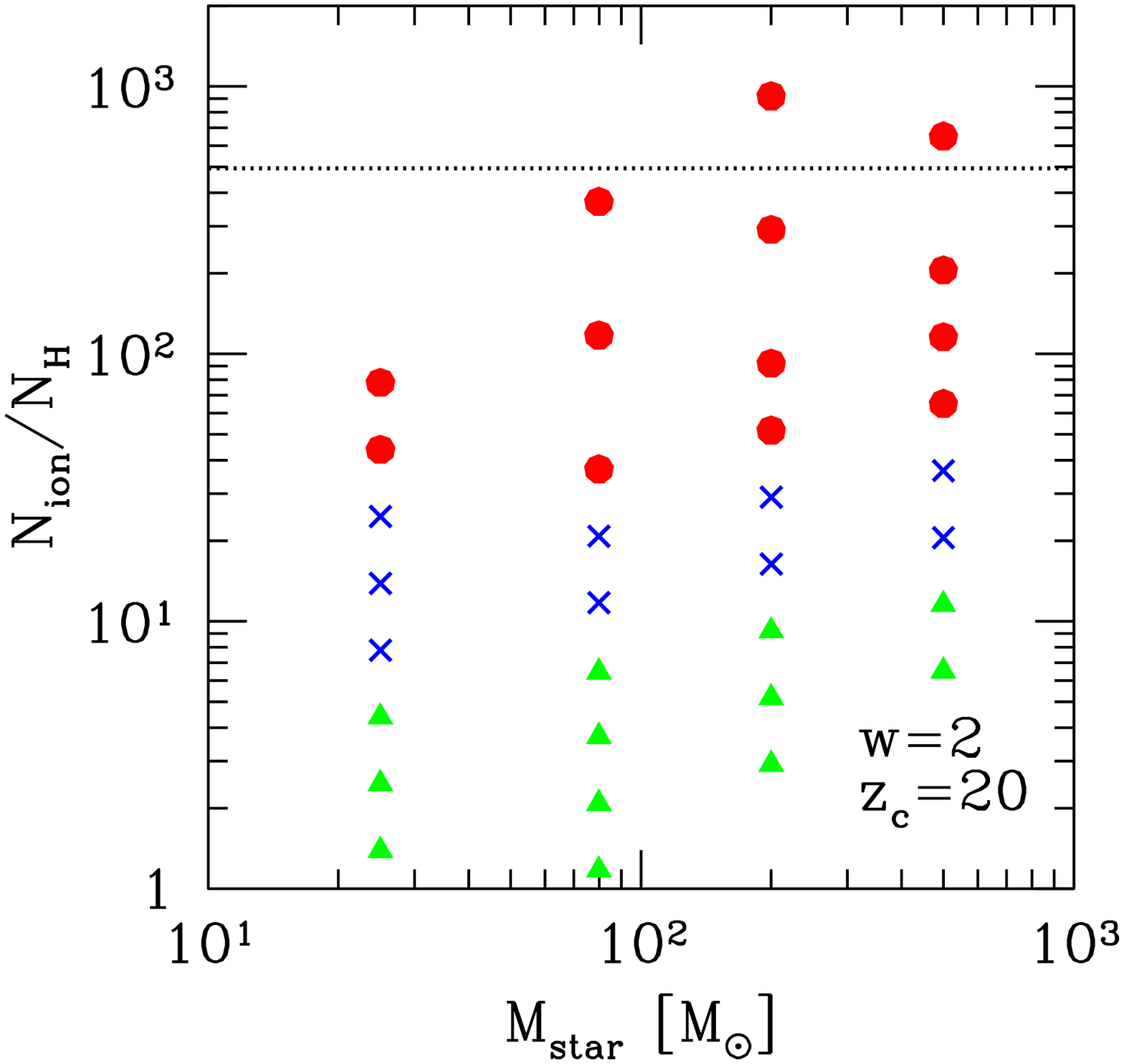} \plotone{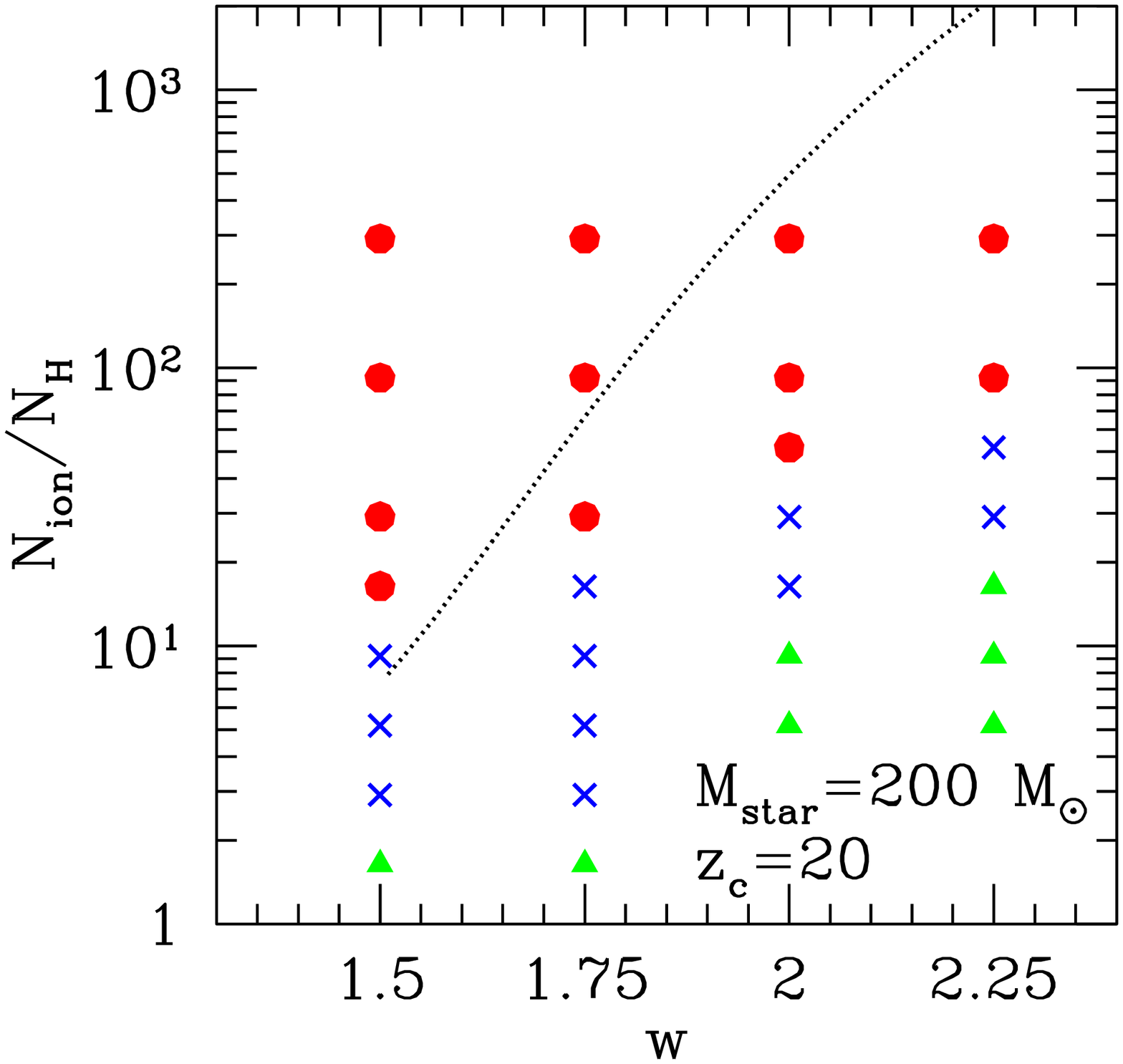}
\plotone{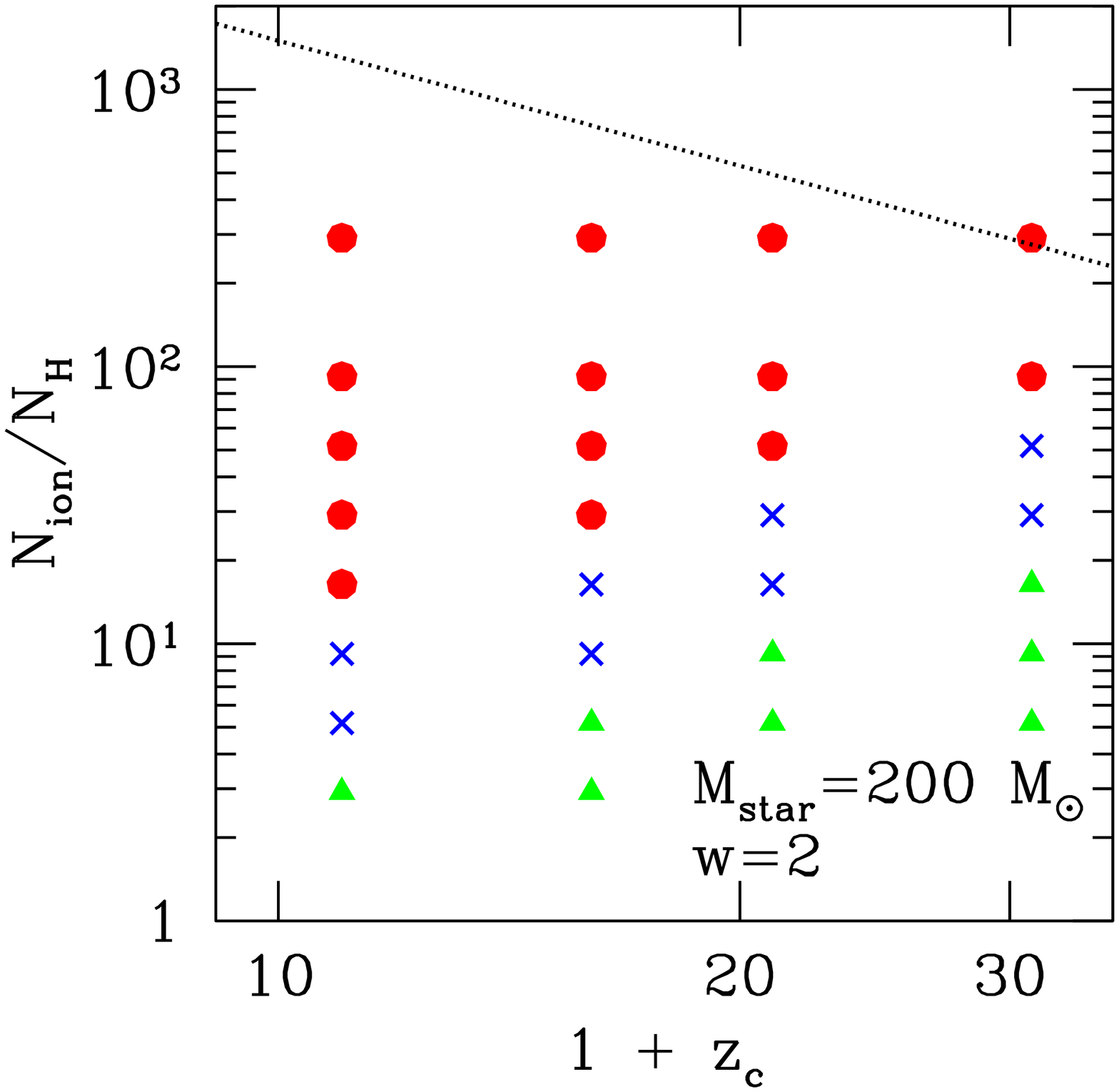} \caption{Total
number of ionizing photons emitted from the central star per hydrogen
atom in a halo as a function of $M_{\rm star}$, $w$, and $z_{\rm
c}$. Meanings of the symbols are the same as in Figure
\ref{fig:msm}. Also plotted for reference by dotted lines are the
effective clumping factor of the halo $C_{\rm halo}$ defined in the main
text.  \label{fig:cms} }
\end{figure}
%%%%%%%%%%%%%%%%%%%%%%%%%%%%%%%%%%%%%%%%%%%%%%

As an alternative way to quantify the strength of radiative feedback, we
plot in Figure \ref{fig:cms} the ratio $N_{\rm ion}/N_{\rm H}$ as a
function of $M_{\rm star}$, $w$, and $z_{\rm c}$, where $N_{\rm ion}$ is
the total number of ionizing photons emitted by the central star and
$N_{\rm H}$ is the total number of hydrogen atoms in a halo.  In order for the
ionizing photons to escape from halos ($f_{\rm esc}^{\rm ion}>1\%$,
circles), $N_{\rm ion}/N_{\rm H} > 10 \sim 100$ is required. The
threshold values of $N_{\rm ion}/N_{\rm H}$ are rather insensitive to
$M_{\rm star}$ but increase with $w$ or $z_{\rm c}$. The H$_2$
photodissociating photons can escape from halos ($f_{\rm esc}^{\rm
LW}>1\%$, crosses and circles) for $N_{\rm ion}/N_{\rm H} > 3 \sim
30$.

It is often assumed in the literature that $N_{\rm ion}/N_{\rm H}$ needs
to be higher than the clumping factor $\langle n^2 \rangle/ \langle n
\rangle^2$ for the gas to be completely ionized. While this assumption
is likely to be valid for nearly uniform media, it cannot be applied to
the present case in which the density gradient is large. To see this
point more clearly, we also display in Figure \ref{fig:cms} the {\it
effective} clumping factor of the halo in its initial configuration for
$3/2 < w < 3$: 
%%%%%%%%%%%%%%%%%%%%%%%%%%%%%%%%%%%%%%%%%%
\begin{eqnarray}
 C_{\rm halo} &=&  \langle n_{\rm i}^2 \rangle_{\rm halo}/\langle n_{\rm i}
\rangle_{\rm halo}^2  \\
&=& \frac{(3-w)^2}{3(2 w-3)} \sbkt{\frac{1}{x^{2w-3}}  -1 }
\frac{1 -x^3}{(1-x^{3-w})^2},  \nonumber 
\end{eqnarray}
%%%%%%%%%%%%%%%%%%%%%%%%%%%%%%%%%%%%%%%%%%
%$C_{\rm halo} = \langle n_{\rm i}^2 \rangle_{\rm halo}/\langle n_{\rm i}
%\rangle_{\rm halo}^2$, 
where $\langle ~ \rangle_{\rm halo}$ denotes the volume average taken at
$r_{\rm in} \leq r \leq r_{\rm vir}$ and $x=r_{\rm in}/r_{\rm vir}$.
The initial condition of the present simulations (eq. [\ref{eq-rin}])
gives 
%%%%%%%%%%%%%%%%%%%%%%%%%%%%%%%%%%%%%%%%%%
\begin{equation}
x = \lbkt{3.93 \times 10^{-7} \sbkt{\frac{1+z_{\rm c}}{20}}^{3}}^{1/w}. 
\label{eq-x}
\end{equation}
%%%%%%%%%%%%%%%%%%%%%%%%%%%%%%%%%%%%%%%%%%
Over the ranges of parameters studied in this paper, $C_{\rm halo}$
increases with increasing $w$ and decreasing $z_{\rm c}$. The dependence
of $C_{\rm halo}$ on $z_{\rm c}$ is simply a product of our choice of
$r_{\rm in}$ (eq. [\ref{eq-rin}]). We have checked that our simulation
results are not sensitive to the specific value of $r_{\rm in}$ (or $x$),
while $C_{\rm halo}$ is.

Figure \ref{fig:cms} shows that the halos are completely ionized and
photons can escape from them even for $N_{\rm ion}/N_{\rm H} < C_{\rm
halo}$. The dependences of $C_{\rm halo}$ on $w$ and $z_{\rm c}$ do not
agree with those of the threshold values of $N_{\rm ion}/N_{\rm H}$ for
$f_{\rm esc}^{\rm ion}>1\%$ either.  These are the consequences of the
enhanced ionization induced by the gas evacuation and the subsequent
R-type I-front propagation. We stress that it is essential to
incorporate the dynamical effects properly, as in equation
(\ref{eq-mcrit}), to account for ionization of the halo with a steep
density gradient.

\subsection{Gas evacuation and the evolution of the 
relic ${\rm H}${\sc ii} regions} 

The \HII regions around a massive short-lived star embedded in a steep
density medium never settle into a static Str\"omgren state. The I-front
keeps expanding outward after the central star turns off.  The relic
\HII regions then start cooling by recombination and inverse-Compton
cooling at such high redshifts, and the ionized fraction rapidly
decreases while the outer part is still expanding. It is therefore
meaningful to investigate how much gas is driven out of a halo in the
end.  In the cosmological context, dark matter halos continue growing,
so it is likely that the expelled gas eventually falls back into the
halo by gravitational force.

\begin{figure}
\plotone{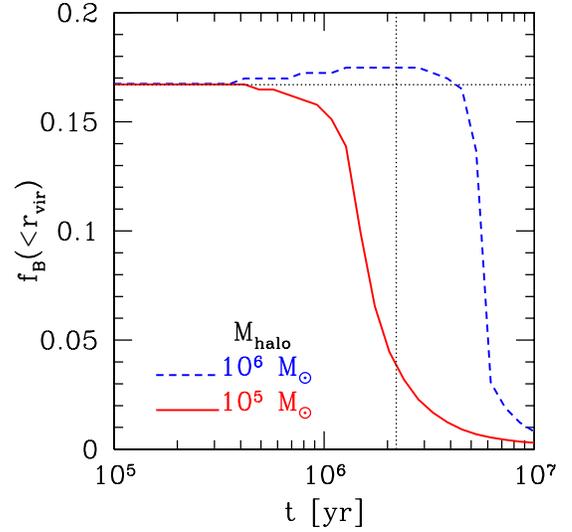} \caption{Evolution of baryon fraction within virial
 radius for $M_{\rm halo}=10^5 \msun$ (solid lines) and $10^6 \msun$
 (dashed lines) in our fiducial runs. Dotted lines denote the lifetime of
 the central star (vertical) and the universal baryon fraction
 $\Omega_{\rm B}/\Omega_{\rm M}$ (horizontal).  \label{fig:fb}}
\end{figure}

Figure \ref{fig:fb} shows the evolution of the baryon fraction within
$r_{\rm vir}$ in the cases of $M_{\rm halo}=10^5$ and $10^6 \msun$ in
our fiducial runs.  The halo gas is almost completely evacuated by $t=2$
Myr for $M_{\rm halo}=10^5 \msun$.  For $M_{\rm halo}=10^6 \msun$, on
the other hand, the gas mass fraction slightly increases by gas
accretion during the first few million years.  This is simply because
the shock radius is $\sim 70$ pc at $t=2$ Myr, still inside the virial
radius (Fig. \ref{fig:evol}).  Even after the central star fades away,
the gas will maintain its outward motion until it dissipates kinetic
energy. The baryon fraction rapidly decreases at $\sim 5$ Myr, when the
shock front overtakes the virial radius.

%\epsscale{1.}
%%%%%%%%%%%%%%%%%%%%%%%%%%%%%%%%%%%%%%%%%%%
%%%%%%%%%%%%%%%%%%%%%%%%%%%%%%%%%%%%%%%%%%%%%%

We have further carried out the runs with the same set of parameters as
those in Figure \ref{fig:fb} but omitting the radiation force term in
equation (\ref{eq:mom}).  While the radiation force enhances the outward
momentum of the gas at small radii (Figures \ref{fig:prof1} and
\ref{fig:prof2}), it helps little to evacuate the gas from a halo; the
baryon fraction within $r_{\rm vir}$ is unchanged to within 5\% when the
radiation force is turned off.

If the central Population III star is so massive that they collapse
directly to form a blackhole (see, e.g. Fryer, Heger \& Woosley 2001),
it does not disturb the gas mechanically any more.  The ultimate fate of
the expelled gas is then described in simple terms.  The cooled gas does
not provide hydrodynamic pressure in the relic \HII region and thus at
least some fraction of the gas eventually falls back toward the halo
center (i.e., gravitational potential well) after a while, gaining
roughly a free-fall velocity.  The free-fall time for a halo collapsing
at $z=15$ is $t_{\rm ff}\sim 50$ Myr in our adopted $\Lambda$CDM
cosmology.  The host halo itself is growing in mass, but the growth
timescale is also of the order $t_{\rm ff}$.  It is thus clear that the
subsequent star-formation is possible only after this fallback of the
gas happens, i.e., gas condensation at the halo center begins only after
a few times $10^7$ yr. Note also that the evolution is more complicated
when the central star explodes as a supernova.  In such cases, the halo
gas may be swept up by the supernova blastwave.

\section{Cosmological Implications}
\label{sec:imp}

\subsection{Suppression of star formation}
\label{sec:sf}

The internal feedback due to photodissociation of hydrogen molecules by
radiation from the very first star likely prohibits further
star-formation within the same halo (Omukai \& Nishi 1999).  Our results
(Figures \ref{fig:fesc_ms}--\ref{fig:fesc_zc}) provide a reasonable
estimation for the critical halo mass that determines the efficiency of
the internal feedback in terms of $f_{\rm esc}^{\rm LW}$.  As shown in
\S 3, early star formation via H$_2$ cooling will be self-regulated
internally and suppressed in halos with $M_{\rm halo} \la 10^7 \msun$.
For such small halos, $f_{\rm esc}^{\rm LW}$ is high and the
'one-star-per-halo' assumption adopted in the present paper is likely to
be adequate.  For halos with sufficiently larger mass and lower $f_{\rm
esc}^{\rm LW}$, the multiple formation of stars or star clusters could
take place.  Therefore, while the very first stars are likely to form in
low-mass halos (Abel, Bryan \& Norman 2002; Bromm, Coppi \& Larson
2002), the subsequent star formation may be suppressed until slightly
larger mass systems begin to collapse. The latter systems may dominate
the cosmic star formation rate and metal production at high redshifts
(e.g. Ricotti et al. 2002).

Note that our results are not restricted by the underlying
one-star-per-halo assumption. We have shown that the emitted number of
ionizing photons, not the number of stars itself, is a key parameter to
control the photon escape fraction (see eq. [\ref{eq-mcrit}]). Given a
weak variation of the main sequence lifetime with mass, $M_{\rm star}$
above $\sim 100 \msun$ can be interpreted as the {\it total mass} of
stars with nearly Eddington luminosities; the case of a single $500
\msun$ star is almost equivalent to that of a few $200 \msun$ stars.
Our results can therefore be rescaled and applied to a wider range of
stellar numbers.

\subsection{Reionization of the Universe}

The present results on the evolution of \HII regions may have profound
implications for reionization of the Universe.  First, high values of
photon escape fractions in low mass halos imply that such halos can be a
major site of photon production at least in the early phase of
reionization.  In order to surpass recombination at $z=20$, the emission
rate of ionizing photons per unit comoving volume must be higher than
$\dot{n}_{\rm ion} = 3.4 \times 10^{51} C_{\rm IGM}$ s$^{-1}$
Mpc$^{-3}$, where $C_{\rm IGM}$ is the clumping factor of the IGM
(eq. [26] of Madau, Haardt \& Rees 1999).  This can be achieved if the
comoving number of halos hosting a $200 \msun$ star is greater than
$13 C_{\rm IGM}$ Mpc$^{-3}$. In the absence of heavy metals, gas can
condense to form stars by H$_2$ cooling only in halos above the mass
$M_{\rm H2} \simeq 5 \times 10^5 h^{-1} \msun$ (Fuller \& Couchman 2000;
Yoshida et al.  2003a). The number of halos just above $M_{\rm H2}$,
expected from the Press \& Schechter (1974) mass function, is $\sim 30$
Mpc$^{-3}$ at $z=20$ in the conventional $\Lambda$CDM model with
$(\Omega_{\rm M}, \Omega_{\Lambda}, h, \Omega_{\rm B}, \sigma_8) = (0.3,
0.7, 0.7, 0.05, 0.84)$. The number of low mass halos is thus comparable
to that needed to ionize the IGM for a moderate value of $C_{\rm
IGM}$. As mentioned below, however, the photon production rate will be
significantly suppressed once secondary feedback effects start to
operate.

Secondly, given the sharp decline in photon escape fractions with
increasing halo mass (Figs \ref{fig:fesc_ms}--\ref{fig:fesc_zc}), there
remains only a narrow range of halo mass in which Population III stars
can form and contribute to reionization.  The photon production can
therefore be stalled readily as the star formation in the low mass halos
is suppressed by, for instance, the rise in the external UV radiation
field (e.g., Machacek, Bryan \& Abel 2001; Kitayama et al. 2001; Yoshida
et al. 2003a; Oh \& Haiman 2003), and the mechanical energy input from
supernovae (e.g., Ricotti \& Ostriker 2004a).  The photon emission may
also decline as the transition from Population III to Population II
stars takes place as a result of metal-enrichment (Cen 2003; Mackey,
Bromm \& Hernquist 2003; Bromm \& Loeb 2003).  Due to high recombination
rate at high redshifts, the ionized fraction may even start decreasing
after a brief period of star-formation in mini-halos.  This can probably
last until a considerable number of stars form in larger systems.  We
will investigate these points more quantitatively in our future
publication.

Finally, evacuation of the halo gas by the central star may further
delay the formation of mini-quasars that are suggested to be alternative
reionization sources in the early universe (e.g. Madau et al. 2004;
Ricotti \& Ostriker 2004b). Such reionization models usually assume
efficient gas accretion onto central blackholes that are the remnants of
very massive ($M_{\rm star}>300\msun$) Population III stars.  Our
calculations, however, indicate that initial gas accretion should be
very inefficient for the blackhole remnants that formed in small mass
($< 10^6 \msun$) halos, because the halo gas is effectively evacuated in
the first place, and, indeed, the gas continues moving outward for over
$10^7$ yr after the central star dies. Similar arguments hold for the
formation of the second generation stars.  It remains to be seen whether
or not early reionization inferred by the WMAP data is achieved in
models with these ionizing sources.

\subsection{Supernova feedback}

Mechanical feedback from the first stars is often cited as a destructive
process in the context of early structure formation.  If the mass of the
central star lies in the range $140 \msun < M_{\rm star} < 260 \msun$,
it will explode as an energetic supernova via pair-instability mechanism
(Barkat, Rakavy, \& Sack 1967; Bond, Arnett, \& Carr 1984; Fryer,
Woosley, \& Heger 2001; Heger \& Woosley 2002).  It is also suggested
that the observed abundance of metal-poor halo stars can be accounted
for by a ``hypernova'' with progenitor mass $20 \msun < M_{\rm star} <
130 \msun$ (Umeda \& Nomoto 2002, 2003).  In either case, the energy
released can be as large as $\sim 10^{53}$ erg, which is much larger
than the gravitational binding energy of a minihalo with mass $\sim 10^6
\msun$.

At the first sight, this simple argument appears to support the notion
that high-$z$ supernovae are enormously destructive.  However, the
evolution of the {\it cooling} supernova remnants (SNRs) crucially
depends on the properties of the surrounding medium, particularly on the
central density after the gas is re-distributed by radiation.  Our
simulations showed that the final gas density at the halo center is
primarily determined by the host halo mass.  For small-mass halos that
are almost completely ionized by a massive star, the central gas density
reduces to $\sim 1$ cm$^{-3}$.  A large fraction of the supernova
explosion energy will then be converted into the kinetic energy of the
ambient media.  On the other hand, for the larger-mass halos, the
central gas density exceeds $10^4$ cm$^{-3}$.  Most of the explosion
energy will be quickly radiated away via free-free emission and the
inverse-Compton scattering of the CMB photons.  The free-free emission
from the hot ($T \sim 10^{6-8}$K) SNR leads to soft-Xray emission.  It
has been suggested that positive feedback, in terms of molecular
hydrogen formation, is possible if an early X-ray background builds up
(Haiman, Abel \& Rees 2001; Oh 2001; Venkatesan, Giroux \& Shull 2001;
see however a counter argument by Machacek et al 2003). Population III
supernova remnants are therefore among the most plausible X-ray sources
in the early universe.  We will study in detail the evolution of the SNR
and the X-ray emission efficiency in a forthcoming paper.

\section{Conclusions}

We have studied the structure and the evolution of early cosmological
\HII regions formed around the first stars. In particular, we addressed
how efficiently the central stars ionize the gas in low-mass halos with
$M_{\rm halo} = 10^5 - 10^8 \msun$ collapsing at $z_{\rm c}=10-30$. We
showed that a single massive star can ionize all the halo gas in a few
million years if the host halo mass is smaller than a few million solar
masses.  The final ionization fraction of the halo gas depends
sensitively on the initial gas density profile as well as the host halo
mass.  While a few previous cosmological simulations suggest an
approximately isothermal density profile ($\rho \propto r^{-2}$) around
the first star-forming regions, the exact slope may vary among the
primordial gas clouds. We therefore carried out a number of simulations
for plausible cases and explored a large parameter space.  For a similar
set of parameters, our results are in good agreement with those of
Whalen et al. (2004), who studied the case of $M_{\rm halo}=7 \times
10^5\msun$ and $z_{\rm c} \sim 20$.

The formation of the \HII region is characterized by initial slow
expansion of a weak D-type ionization front near the center, followed by
rapid propagation of an R-type front throughout the outer gas envelope.
We find that the transition between the two front types is indeed a
critical condition for the complete ionization of halos of cosmological
interest.  This accounts for the fact that the photon escape fraction
has a sharp transition from $\sim 100\%$ to $\sim 0 \%$ at a certain
critical mass scale of $10^6 \sim 10^7 \msun$. It is also responsible for the
fact that the whole halo can be ionized under smaller values of
photon-to-baryon ratio than the effective clumping factor.  The
radiation force can contribute to enhancing the outward motion within
$\sim 30$ pc at the initial stage of expansion.

Based on the obtained numerical results, we developed an analytic model
to predict the escape fractions of ionizing photons and dissociating
photons. We find that there is a narrow range of halo mass in which
Population III stars can form and contribute to ionizing the Universe,
provided that low mass halos are in infall phases and host a small
number of massive stars. The analytic model and the numerical results
presented in this paper provide useful ingredients for the studies of
early reionization, the evolution of high-redshift supernova remnants,
and the formation of the second generation stars.

\hfill 

We thank Tom Abel, S. Peng Oh, and Kazuyuki Omukai for fruitful
discussions, and the referee for useful comments. NY acknowledges
support from the Japan Society of Promotion of Science Special Research
Fellowship (02674).  This work is supported in part by the Grants-in-Aid
by the Ministry of Education, Science and Culture of Japan (14740133:TK,
15740122:HS, 15340060:MU).

%\clearpage

\appendix
\section{Implementation of Radiative Transfer}

Above the Lyman limit of hydrogen ($h\nu \geq 13.6$ eV), we solve the
following radiative transfer equation taking account of both absorption
and emission:
%%%%%%%%%%%%%%%%%%%%%%%%%%%%%%
\begin{equation}
\frac{d I_\nu}{d s} = - n_{\rm HI} \sigma_\nu I_\nu + j_\nu, 
\label{eq-transf} 
\end{equation}
%%%%%%%%%%%%%%%%%%%%%%%%%%%%%%
where $I_\nu$ is the specific intensity at frequency $\nu$, $s$ is the
distance along a light ray, $\sigma_\nu$ is the photoionization cross
section of hydrogen (Osterbrock 1989), and $j_\nu$ is the local emission
coefficient for ionizing radiation.  In the case of $kT \ll h\nu_{\rm L}$ as in
the present simulations, $j_\nu$ is given by (Tajiri \& Umemura 1998;
Susa \& Umemura 2000)
%%%%%%%%%%%%%%%%%%%%%%%%%%%%%%
\begin{equation}
j_\nu =  
\left\{\begin{array}{ll}
   \frac{h \nu \alpha_1 n_{\rm e} n_{\rm HII}}
{4 \pi \delta \nu} & \nu_{\rm L} \leq
\nu < \nu_{\rm L} + \delta \nu, \\
0 & \nu \geq \nu_{\rm L} + \delta \nu, \\
 \end{array} \right.  
\label{eq-source} 
\end{equation}
%%%%%%%%%%%%%%%%%%%%%%%%%%%%%%
where $\nu_{\rm L}$ is the Lyman limit frequency, $\alpha_1$ is the
recombination rate to the ground state of hydrogen (Osterbrock 1989),
and $\delta \nu = kT/h$ is the thermal width of the recombination line.
Recombinations to the excited states are excluded in equation
(\ref{eq-source}) because they are unable to produce ionizing photons
above the Lyman limit. In practice, equation (\ref{eq-transf}) is solved
separately for $\nu_{\rm L} \leq \nu < \nu_{\rm L} + \delta \nu$ and
$\nu \geq \nu_{\rm L} + \delta \nu$. For the latter, photoionization is
regarded as pure absorption and the solution reduces to
%%%%%%%%%%%%%%%%%%%%%%%%%%%%%%
\begin{equation}
 I_\nu(s) = I_\nu(0) ~\exp \sbkt{- \sigma_\nu {\cal N}_{\rm HI}}, 
~~~ (\nu \geq \nu_{\rm L} + \delta \nu)
\label{eq-sol}
\end{equation}
%%%%%%%%%%%%%%%%%
where ${\cal N}_{\rm HI}= \int_0^s n_{\rm HI} ds'$ is the column density
of neutral hydrogen along the photon path.

The coefficient (per unit time) and the heating rate (energy per unit
time per unit volume) for photoionization of hydrogen atoms (${\rm H} +
h\nu \rightarrow e^- + {\rm H}^+$) are then given respectively by
%%%%%%%%%%%%%%%%%%%%%%%%%%%%%%
\begin{eqnarray}
\label{eq-gamma}
k_{\rm ion} &=& \int d\Omega \int_{\nu_{\rm L}}^\infty d\nu 
\frac{I_\nu \sigma_\nu }{h \nu}, \\
{\cal H} &=& n_{\rm HI} \int d\Omega \int_{\nu_{\rm L}}^\infty d\nu 
\frac{I_\nu \sigma_\nu }
{h \nu} (h \nu - h \nu_{\rm _L}),  
\label{eq-heat}
\end{eqnarray}
%%%%%%%%%%%%%%%%%%%%%%%%%%%%%%
where the integral with respect to $\Omega$ is taken over
all solid angles. Assuming that the momentum of a photon is transfered
to the gas upon each ionization, the radiation force per unit mass in
the direction specified by a unit vector $\vec{n}$ is expressed as
%%%%%%%%%%%%%%%%%%%%%%%%%%%%%%
\begin{equation}
f_{\rm rad}(\vec{n}) = \frac{X_{\rm HI}}{m_{\rm p} c}  
\int d\Omega \int_{\nu_{\rm L}}^\infty d\nu 
~ I_\nu \sigma_\nu \cos \theta 
\label{eq-frad}
\end{equation}
%%%%%%%%%%%%%%%%%%%%%%%%%%%%%%
where $\theta$ is the angle between $\vec{n}$ and the light ray, $m_{\rm
p}$ is the proton mass, and $c$ is the speed of light. In our
spherically symmetric simulations, $\vec{n}$ lies in the radial
direction. Photon momentum carried away by reprocessed radiation is
properly taken into account by using the solutions of multi-frequency
radiative transfer mentioned above.  

As far as only the radiation from a central star is concerned, the
intensity integrated over solid angles is well approximated at $\nu \geq
\nu_{\rm L} + \delta \nu$ by
%%%%%%%%%%%%%%%%%%%%%%%%%%%%%%
\begin{equation}
\int I_\nu d\Omega \simeq \int I_\nu \cos \theta d\Omega = 
\frac{L_\nu \exp(- \sigma_\nu {\cal N}_{\rm HI})}{4 \pi r^2}, 
\end{equation}
%%%%%%%%%%%%%%%%%%%%%%%%%%%%%%
where $L_\nu$ is the stellar luminosity, and $r$ is the radial distance
from the center, taken to be sufficiently larger than the physical size
of the star.  
%The computational time of equations (\ref{eq-gamma})--(\ref{eq-frad})
%can further be reduced by preparing numerical tables for $\int_{\nu_{\rm
%L}}^\infty d\nu L_\nu \sigma_\nu \exp(- \sigma_\nu {\cal N}_{\rm HI}) $
%and $\int_{\nu_{\rm L}}^\infty (d\nu/\nu) L_\nu \sigma_\nu \exp(-
%\sigma_\nu {\cal N}_{\rm HI})$ as a function of ${\cal N}_{\rm HI}$.
On the other hand, the diffuse radiation just above the Lyman limit,
$\nu_{\rm L} \leq \nu < \nu_{\rm L} + \delta \nu$, comes from all the
directions.  For the diffuse component, we directly solve equation
(\ref{eq-transf}) by means of an impact parameter method (Mihalas \&
Weibel-Mihalas 1984).  At each radial point, angular integrations in
equations (\ref{eq-gamma})--(\ref{eq-frad}) are done over at least 20
bins in $\theta = 0 -\pi$. This is achieved by handling 400--1,000
impact parameters for light rays.

In the LW (11.2--13.6 eV) band, we use the self-shielding function of
Draine \& Bertoldi (1996) to compute the coefficient (per unit time) for
photodissociation of H$_2$ molecules (${\rm H}_2 + h\nu \rightarrow {\rm
H}+ {\rm H}$):
%%%%%%%%%%%%%%%%%%%%%%%%%%%%%%
\begin{equation}
 k_{\rm H2} =  0.15 \xi_{\rm pump} \chi f_{\rm shield},
\end{equation}
%%%%%%%%%%%%%%%%%%%%%%%%%%%%%%
where 
%%%%%%%%%%%%%%%%%%%%%%%%%%%%%%
\begin{eqnarray}
 \xi_{\rm pump} &\simeq& 3 \times 10^{-10} \mbox{ s$^{-1}$},  \\
  \chi &=& \frac{\lbkt{\nu/c \int I_\nu d\Omega }_{\nu/c = 1000 \AA}}
{4 \times 10^{-14} \mbox{ erg cm$^{-3}$}} \simeq 
\frac{\lbkt{\nu L_\nu /(4 \pi c r^2)}_{\nu/c = 1000 \AA}}
{4 \times 10^{-14} \mbox{ erg cm$^{-3}$}} , \\
 f_{\rm shield} &=& \min \lbkt{1, 
\sbkt{\frac{{\cal N}_{\rm H2}}{10^{14} \mbox{ cm}^{-2}}}^{-0.75}},
\end{eqnarray}
%%%%%%%%%%%%%%%%%%%%%%%%%%%%%%
and ${\cal N}_{\rm H2}$ is the column density of H$_2$.

We also compute the coefficients for photodetachment of H$^-$ (${\rm
H}^- + h\nu \rightarrow e^- + {\rm H}$) and photodissociation of H$_2^+$
(${\rm H}_2^+ + h\nu \rightarrow {\rm H}^+ + {\rm H}$) from the
radiation below $\nu_{\rm L}$ as
%%%%%%%%%%%%%%%%%%%%%%%%%%%%%%
\begin{eqnarray}
 k_{\rm i} =  \int d\Omega \int_{\nu_{\rm i}}^{\nu_{\rm L}} d\nu 
\frac{I_\nu \sigma_{{\rm i}, \nu}}{h \nu} \simeq \frac{1}{4 \pi r^2}  \int_{\nu_{\rm i}}^{\nu_{\rm L}} d\nu 
\frac{L_\nu \sigma_{{\rm i}, \nu}}{h \nu} ,
\end{eqnarray}
%%%%%%%%%%%%%%%%%%%%%%%%%%%%%%
where the subscript i stands for H$^-$ or H$_2^+$ and $\sigma_{{\rm
i},\nu}$ denotes the cross section. For photodetachment of H$^-$, we use
a fitting formula by Tegmark et al. (1997) to the cross section of
Wishart (1979) with the low-energy cut-off $h \nu_{{\rm H}^-} = 0.74$
eV. For photodissociation of H$_2^+$, we use the cross section from
Table 2 of Stancil (1994) down to $h \nu_{{\rm H}_2^+} = 0.062$ eV. Our
results are insensitive to specific values of the low-energy cut-off,
because black-body spectra of stellar sources diminish at low
energies. The radiation above $\nu_{\rm L}$ is also irrelevant since it
will be heavily absorbed by hydrogen atoms whenever H$^-$ and H$_2^+$
processes become relevant.

\section{Code Test}

We have tested and verified the accuracy of our Lagrangian code by a
variety of test problems, such as the shock-tube problem, the Sedov
explosion problem (Sedov 1959), and a comparison with the self-similar
solution for the adiabatic accretion of collisional gas (Bertschinger
1985). We here describe the results of a ``Str\"omgren sphere'' test,
which is the most relevant to the present paper.

\begin{figure}
\plotone{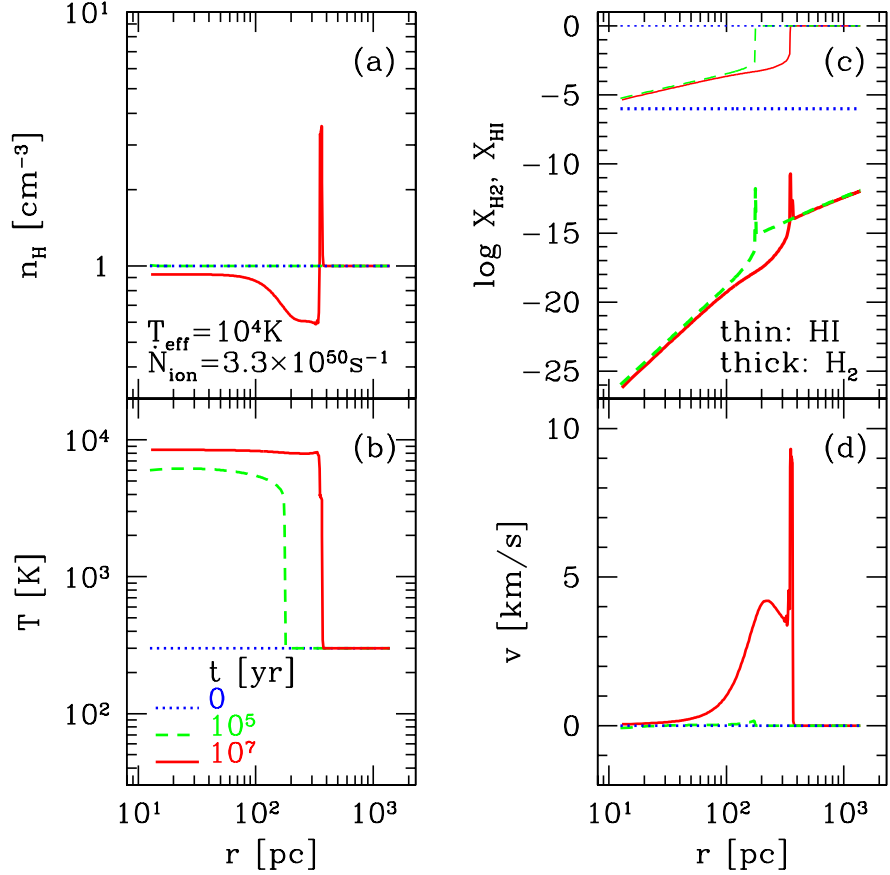} \caption{Propagation of an I-front in a uniform medium
 around a source with $\dot{N}_{\rm ion} = 3.3 \times 10^{50}$ s$^{-1}$
 and $T_{\rm eff}=10^4$ K. Radial profiles are shown at $t=0$ (dotted
 lines), $10^5$ yr (dashed lines), and $10^7$ yr (solid lines) for (a)
 hydrogen density, (b) temperature, (c) HI (thin lines) and H$_2$ (thick
 lines) fractions, and (f) radial velocity.  The Str\"omgren radius is
 at $r_{\rm St}=220$ pc.  \label{fig:strom1}}
\end{figure}

We simulate the propagation of an I-front around a central
source with $\dot{N}_{\rm ion} = 3.3 \times 10^{50}$ s$^{-1}$ and
$T_{\rm eff}=10^4$ K in a uniform medium with $n_{\rm H}=1$ cm$^{-3}$.
The Str\"omgren radius is at $r_{\rm St}=220$ pc for $T=10^4$ K.  For
the sake of direct comparison with analytic solutions, we exclude the
radiation force and gravitational force terms from the momentum equation
(eq. [\ref{eq:mom}]).  Figure \ref{fig:strom1} shows the radial profiles
of hydrogen density, temperature, HI and H$_2$ fractions, and the gas
velocity. In contrast to the cases with steep density gradient
($w>3/2$), the I-front is initially R-type at $r < r_{\rm St}$ and then
changes into D-type at $r > r_{\rm St}$. The I-front is resolved with
more than 10 gas shells.  As in Figure \ref{fig:prof1}, there temporally
appears a thin shell of H$_2$ just in front of the \HII region.

\begin{figure}
\plotone{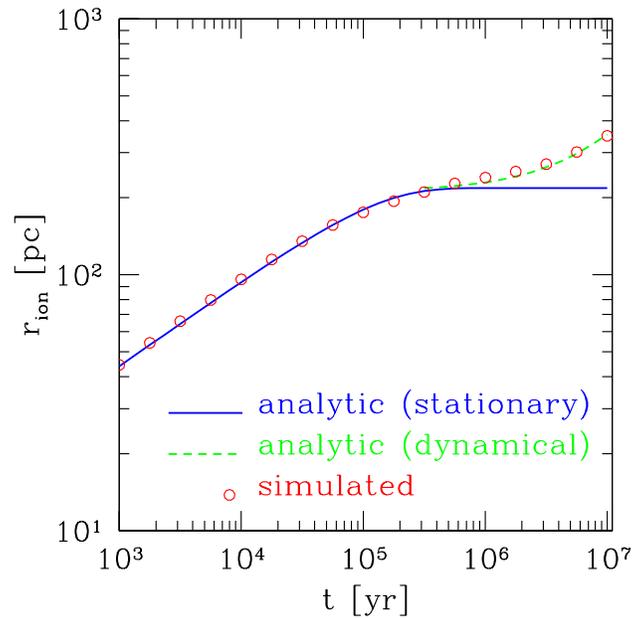} \caption{Evolution of an I-front in the
simulation shown in Figure \ref{fig:strom1} (circles). Lines denote 
the analytic solutions for the I-front propagation in stationary (solid)
and dynamical (dashed) media described in the text.  
\label{fig:strom2}
}
\end{figure}

In Figure \ref{fig:strom2}, we also plot the evolution of the I-front
radius. For definiteness, we define it as the radius at which HI
fraction is equal to 0.1. Also plotted for reference are the analytic
solutions of I-front propagation in stationary and dynamical media (eqs
[12-9] and [12-20] of Spitzer 1978) for the photoheated temperature of
$T=10^4$ K. They are expected to describe the evolution of R-type and
D-type fronts, respectively. Our simulation results in fact reproduce
both of them within 5\% accuracy at $t > 10^3$ yr. This further ensures
the capability of the code for the simulations in the present paper.

\epsscale{.8}
%%%%%%%%%%%%%%%%%%%%%%%%%%%%%%%%%%%%%%%%%%%
%%%%%%%%%%%%%%%%%%%%%%%%%%%%%%%%%%%%%%%%%%%%%%
\epsscale{.55}
%%%%%%%%%%%%%%%%%%%%%%%%%%%%%%%%%%%%%%%%%%%%%%
%%%%%%%%%%%%%%%%%%%%%%%%%%%%%%%%%%%%%%%%%%%%%
\end{document}